\documentclass[a4paper,usenatbib]{mnras}
\usepackage{aas_macros}
\usepackage{ae,aecompl}
\usepackage{graphicx}
\usepackage{amsmath}
\usepackage{amssymb}
\usepackage[dvipsnames]{xcolor}
\hypersetup{colorlinks=true,allcolors=teal}

\newcommand{\Mh}{\ensuremath{h^{-1}M_{\odot}}}

\newcommand{\Mpch}{\ensuremath{h^{-1}{\rm Mpc}}}

\newcommand{\avg}[1]{\ensuremath{\left\langle \,#1\, \right\rangle}}

\newcommand{\der}{\ensuremath{{\rm d}}}

\newcommand{\erf}[1]{\ensuremath{{\rm erf}\left(#1\right)}}

\newcommand{\eqn}[1]{equation~\eqref{#1}}
\newcommand{\eqns}[1]{equations~\eqref{#1}}

\newcommand{\ph}[1]{\phantom{#1}}

\newcommand{\be}{\begin{equation}}
\newcommand{\ee}{\end{equation}}
\newcommand{\Cal}[1]{\ensuremath{\mathcal{#1}}}

\newcommand{\ngal}{\ensuremath{\bar n_{\rm g}}}
\newcommand{\Ns}{\ensuremath{\bar N_{\rm sat}}}
\newcommand{\cNs}{\ensuremath{\Cal{N}_{\rm sat}}}
\newcommand{\fcen}{\ensuremath{f_{\rm cen}}}
\newcommand{\bbar}{\ensuremath{\bar b}}
\newcommand{\nrgal}{\ensuremath{\bar n_{\rm r}}}

\newcommand{\Nrs}{\ensuremath{\bar N_{\rm rsat}}}
\newcommand{\cNrs}{\ensuremath{\Cal{N}_{\rm rsat}}}
\newcommand{\frcen}{\ensuremath{f_{\rm rcen}}}
\newcommand{\brbar}{\ensuremath{\bar b_{\rm r}}}

\newcommand{\nbgal}{\ensuremath{\bar n_{\rm b}}}

\newcommand{\cNbs}{\ensuremath{\Cal{N}_{\rm bsat}}}
\newcommand{\fbcen}{\ensuremath{f_{\rm bcen}}}
\newcommand{\bbbar}{\ensuremath{\bar b_{\rm b}}}

\title[Analytical halo model of galactic conformity]
  {Analytical halo model of galactic conformity}
\author[Pahwa \& Paranjape]{
Isha Pahwa,$^{1,2}$\thanks{E-mail: ipahwa@iucaa.in}
and Aseem Paranjape$^{1}$\thanks{E-mail: aseem@iucaa.in}
\\
$^{1}$Inter-University Centre for Astronomy and Astrophysics, Ganeshkhind, Post Bag 4, Pune 411007, India\\
$^{2}$Leibniz-Institut f\"ur Astrophysik Potsdam (AIP), An der Sternwarte 16, D-14482 Potsdam, Germany\\
}

\date{draft}

\pagerange{\pageref{firstpage}--\pageref{lastpage}} \pubyear{2016}

\begin{document}

\label{firstpage}

\maketitle

\begin{abstract}
\noindent
We present a fully analytical halo model of colour-dependent clustering that incorporates the effects of galactic conformity in a halo occupation distribution (HOD) framework. 
The model, based on our previous numerical work, describes conformity through a correlation between the colour of a galaxy and the concentration of its parent halo, leading to a correlation between central and satellite galaxy colours at fixed halo mass. 
The strength of the correlation is set by a tunable `group quenching efficiency', and the model can separately describe group-level correlations between galaxy colour (1-halo conformity) and large scale correlations induced by assembly bias (2-halo conformity). 
We validate our analytical results using clustering measurements in mock galaxy catalogs, finding that the model is accurate at the 10-20 percent level for a wide range of luminosities and length scales. 
We apply the formalism to interpret the colour-dependent clustering of galaxies in the Sloan Digital Sky Survey (SDSS). 
We find good overall agreement between the data and a model that has 1-halo conformity at a level consistent with previous results based on an SDSS group catalog, although the clustering data require satellites to be redder than suggested by the group catalog. 
Within our modelling uncertainties, however, we do not find strong evidence of 2-halo conformity driven by assembly bias in SDSS clustering. 
\end{abstract}

 
\section{Introduction}
\label{sec:intro}
\noindent
The Halo Model for galaxies \citep[see][for a review]{cs02} posits that all galaxies occupy and evolve in the gravitational potential wells of virialised dark matter haloes. Observables such as the luminosity function, stellar mass function and spatial clustering of galaxies can then be thought of as, simply, averages over the mass function and clustering of \emph{haloes}, weighted by the number of galaxies of a given type that reside in haloes of a given type \citep{seljak00,sshj01,bw02}. The simplest of this class of models -- in which the number of galaxies of a certain luminosity or stellar mass in a halo is determined solely by the mass of the halo -- has found considerable observational support in recent years \citep{zehavi+05,zehavi+2011,guo+15}. Observations also indicate that galaxies in groups fall into two populations with somewhat different properties, `centrals' which typically reside close to the gravitational center of the group's parent halo (and are usually the brightest object in the group) and `satellites' which orbit the central. This is incorporated in the Halo Model by separately modelling the halo occupancy functions of centrals and satellites \citep{berlind+03,yang+03,zheng04,vo04,zehavi+05,zehavi+2011,guo+14,guo+16}. 

The result is a convenient and flexible statistical model, in which the physics of galaxy formation and evolution is absorbed into parametrised functions describing the number of galaxies in haloes of mass $m$. When combined with an observationally constrained prescription for assigning colours to galaxies \citep[][hereafter, S09]{ss09}, this model can also explain the observed differences between the abundances and clustering of red and blue galaxies at fixed luminosity \citep[][hereafter, Z11]{zehavi+2011}. At the faint end, for example, blue galaxies in the S09 model are predominantly centrals in low-mass haloes, while a large fraction of red galaxies of the same luminosity are satellites in massive haloes. This is, of course, in line with the expectation that satellites in massive haloes may be subjected to several physical processes that could lead to the cessation of star formation \citep[see][for a review]{sd15}.
Since halo mass correlates strongly with environment, this also amounts to explaining well-established observational trends such as the colour-density scaling \citep{lewis+02,kauffmann+04,muldrew+12}. 

This simple `mass-only' Halo Model of galaxies fails, however, in explaining galactic conformity, which is the observation that satellites in groups with star forming centrals are preferentially star forming, \emph{even at fixed halo mass}. Originally noticed (and defined) by \citet{weinmann+06a} using the group catalog of \citet{yang+05}, this effect has been confirmed by several authors since \citep{wang+10,wang+12,kauffmann+10,robotham+13,phillips+13,knobel+15,pkhp15}. Modelling this effect would necessarily involve introducing a dependence of star formation properties on one or more variables \emph{in addition to} halo mass.

Intriguingly, a similar effect was observed by \citet{kauffmann+13} at very large spatial separations ($\sim4$Mpc in projection), leading to a discussion in the literature on the potential causes for a large scale, or `2-halo', correlation, in contrast to the `1-halo' conformity of \citet{weinmann+06a} which could presumably be explained by group-scale physical processes \citep{hearin+15a}. The primary suspect for large scale correlations between star formation properties of galaxies at fixed halo mass is the correlation of halo assembly history with environment, or `halo assembly bias' \citep{st04,gsw05,wechsler+06,jing+07,dalal+08,desjacques08a,hahn+09,fm10}. A corresponding \emph{galaxy} assembly bias has been challenging to establish observationally \citep{lin+16,miyatake+16,saito+16,twcm16,zu+17}; however, with galactic conformity potentially a smoking gun for assembly bias, it becomes interesting to develop models that can distinguish group-scale effects from those of assembly bias.
This has led to recent efforts in developing abundance matching models that take into account galactic conformity by correlating galaxy colours with the formation histories of (sub)haloes in high resolution $N$-body simulations \citep{hw13,masaki+13,hearin+14a}.

Recently, \citet[][hereafter, P15]{pkhp15}, presented an extension of the Halo Occupation Distribution (HOD) framework -- in particular, the colour HOD model of S09 mentioned earlier -- to include a simple and tunable prescription for galactic conformity. This was accomplished by correlating the colour of a galaxy with the \emph{concentration} of its parent halo. The choice of halo concentration as the additional parameter controlling galaxy properties was motivated by the fact that the distribution of this variable shows strong environmental trends \emph{at fixed halo mass} \citep{wechsler+06,jing+07,fw10}, in addition to being tightly correlated with halo age \citep{navarro+97,wechsler+02}. P15 showed that this model is flexible enough that one can switch on and off the effects of large scale assembly bias on galaxy colours (by scrambling concentrations at fixed halo mass) while keeping 1-halo conformity intact; this was used to show that group-scale effects could easily extend to several Mpc if one were not careful in selecting samples by halo mass. Later work has also generalised this theme to include variables other than halo concentration in modelling assembly bias \citep{hearin+16}.

In this paper, we present the analytical framework for galaxy clustering in the conformity model of P15. As we will demonstrate, this essentially amounts to accounting for both halo mass and halo concentration, by replacing the halo mass function $n(m)$ with the joint distribution $n(m,c)$ of mass and concentration, and correspondingly tracking the concentration dependence of the modified HOD from P15. This is similar in spirit to the recently introduced `decorated HOD' framework implemented using \textsc{Halotools} \citep{hearin+16}, which is a set of general purpose semi-numerical tools for exploring the effects of assembly bias using high resolution $N$-body simulations. Our approach complements such efforts, in that it is fully analytical and therefore does not rely on access to large simulations.  We will validate our analytical results using the mock catalogs described by P15, showing that our model for colour-dependent clustering is accurate to within $\sim10$-$20\%$.

To demonstrate the utility of the formalism, we will compare our analytical results with the clustering measurements of galaxies in the Sloan Digital Sky Survey\footnote{http://www.sdss.org} \citep[SDSS][]{york+00} presented by Z11, exploring in particular the degeneracy between the level of galactic conformity and the satellite red fraction. Although our formalism is quite general and is equally applicable for modelling galaxies split by luminosity or stellar mass, we will use the HOD calibration of Z11 -- who segregated galaxies by luminosity -- as the basis of the results we display. Consequently, in this work we will focus on modelling observables as a function of luminosity rather than stellar mass.

The paper is organised as follows. In section~\ref{sec:global}, we recapitulate the standard HOD framework (this will set our notation) and show how to incorporate the P15 model. We describe the modified correlation functions in section~\ref{sec:corrfunc}. We validate our results against mock catalogs in section~\ref{sec:results}, followed by an application of our model to SDSS measurements in section~\ref{sec:appl}. We conclude in section~\ref{sec:concl}. Throughout, we will denote halo masses $m$ in units of $h^{-1} M_{\odot}$, where $M_{\odot}$ is the mass of sun and $H_0 = 100 h$ km s$^{-1}$ Mpc$^{-1}$ is the Hubble constant. We use a flat $\Lambda$-cold dark matter ($\Lambda$CDM) cosmology with parameters $\Omega_m = 0.25, \, \Omega_b=0.045, \, h = 0.7, \, \sigma_8 = 0.8$ and $n_s = 0.96$, which are consistent with the 5 year results of the WMAP experiment~\citep{hinshaw+09}, as well as with the cosmology used by Z11 in calibrating their HOD.

\section{HOD framework}
\label{sec:global}
\noindent
In this section, we recapitulate the standard HOD framework for populating galaxies in haloes, and describe its extension to incorporate the P15 conformity model.

\subsection{Standard HOD Modelling}
\label{sec:global:subsec:HOD}
\noindent
In the standard approach, the abundances of central and satellite galaxies are modelled separately. Let $\fcen(>L|m)$ be the fraction of $m$-haloes (i.e., haloes with masses in the range $(m,m+\der m)$) that have a central galaxy brighter than the luminosity threshold $L$. The number of satellites brighter than $L$ in each $m$-halo with a central brighter than $L$ is assumed to be Poisson distributed with mean $\Ns(>L|m)$. If a halo does not contain a central brighter than $L$, it is assumed to have no satellites brighter than $L$ either. The luminosities of centrals and satellites then follow the distributions $\fcen(>L|m)/\fcen(>L_{\rm min}|m)$ and $\Ns(>L|m)/\Ns(>L_{\rm min}|m)$, respectively, where $L_{\rm min}$ is some chosen threshold that defines the luminosity-complete sample.

The functions $\fcen(>L|m)$ and $\Ns(>L|m)$ define the HOD, with the mean number of galaxies brighter than $L$ residing in $m$-haloes given by $\bar N_{\rm gal}(>L|m) = \fcen(>L|m)\left[1+\Ns(>L|m)\right]$. In this work, we will use the forms calibrated by Z11: 
\begin{align}
\fcen(>L|m) &= \frac12\left[1+\erf{\frac{\log(m/M_{\rm min})}{\sigma_{\log M}}}\right]\,,
\label{hod-fcen}\\
\Ns(>L|m) &= \left(\frac{m-M_0}{M_1}\right)^\alpha\,,
\label{hod-Nsat}
\end{align}
with $\{M_{\rm min},\sigma_{\log M},\alpha,M_1,M_0\}$ being functions of the threshold $L$, for which we use an interpolation kindly provided by Ramin Skibba \citep[see also Appendix A2 of][]{ss09}. 

We define the following quantities that are fixed by the HOD:
\begin{align}
\cNs(>L|m) &\equiv \fcen(>L|m)\Ns(>L|m)\notag\\
\cNs(L|m) &\equiv -\partial \cNs(>L|m) / \partial L\notag\\
\fcen(L|m) &\equiv -\partial \fcen(>L|m) / \partial L\,,\notag\\
\Ns(L|m) &\equiv \cNs(L|m) / \fcen(>L|m)\,.
\label{HODderivs}
\end{align}
Here $\fcen(L|m)\der L$ is the differential fraction of $m$-haloes occupied by $L$-centrals (the terminology being analogous to $m$-haloes). Similarly, $\Ns(L|m)\der L$ is the mean differential number of $L$-satellites in $m$-haloes whose central is brighter than $L$. 

Let $p({\rm sat}|L,m)$ be the fraction of $L$-galaxies that are satellites, with the corresponding fraction for centrals given by $p({\rm cen}|L,m)=1-p({\rm sat}|L,m)$. The satellite fraction is fixed by the HOD as:
\be
p({\rm sat}|L,m) = \cNs(L|m)/\left(\fcen(L|m)+\cNs(L|m)\right)\,.
\label{psat-massdep}
\ee
In practice, one usually deals with absolute magnitudes rather than luminosities, in which case the luminosity `$L$' should be replaced with, in our case, the SDSS $r$-band absolute magnitude `$M_r$' in the results that follow; e.g., $\fcen(>L|m) = \fcen(<M_r|m)$, $\fcen(L|m)\der L = \fcen(M_r|m)\der M_r$.

\subsection{Colour-selected HOD}
\label{sec:global:subsec:redfrac}
\noindent
The basic S09 Halo Model for colours works with conditional colour distributions at fixed galaxy luminosity and halo mass. The bimodality of colours observed in SDSS \citep{baldry+04} makes it convenient to further split galaxies into `red' and `blue' based on, e.g., whether their $g-r$ colours are above or below some threshold  \citep[][S09; Z11]{vdb+08}. The \emph{modelled} red fraction of $L$-galaxies in $m$-haloes, $p({\rm red}|L,m)$, can then be written as
\begin{align}
p({\rm red}|L,m) &= p({\rm red}|{\rm cen},L)\,p({\rm cen}|L,m)\notag\\ 
&\ph{p({\rm red}|{\rm cen},L)} + p({\rm red}|{\rm sat},L)\,p({\rm sat}|L,m)\,,
\label{cen-sat-split-S09}
\end{align}
where $p({\rm red}|{\rm sat/cen},L)$ gives the fraction of satellites/centrals that are red. As discussed in detail by S09 and P15, it is a reasonable approximation to ignore any halo mass dependence in these red fractions, even in the presence of conformity. These fractions are, however, constrained by the fact that the halo-averaged version of \eqn{cen-sat-split-S09} must return the \emph{measured} all-galaxy red fraction:
\begin{align}
p({\rm red}|L) &= p({\rm red}|{\rm cen},L)\,\bar p({\rm cen}|L) \notag\\
&\ph{p({\rm red}|{\rm cen})}
+ p({\rm red}|{\rm sat},L)\,\bar p({\rm sat}|L)\,,
\label{cen-sat-split-avg}
\end{align}
where
\begin{align}
&\bar p({\rm sat}|L) = \frac{\int\der m\,n(m)\,\cNs(L|m)}{\int\der m\,n(m)\,\left[\fcen(L|m)+\cNs(L|m)\right]}\,,
\label{pbarsat}
\end{align}
and $\bar p({\rm cen}|L) = 1-\bar p({\rm sat}|L)$, with $n(m)\der m$ being the number density of $m$-haloes (the halo mass function)\footnote{Throughout, we use the fitting formula of \citet{Tinker08} for calculating the halo mass function using the $m_{\rm 200b}$ mass definition. Here $m_{\rm 200b}$ is the mass contained in a radius $R_{\rm 200b}$ around the halo center-of-mass at which the enclosed dark matter density becomes $200$ times the mean density of the Universe.}.

\begin{figure}
\centering
\includegraphics[width=0.45\textwidth]{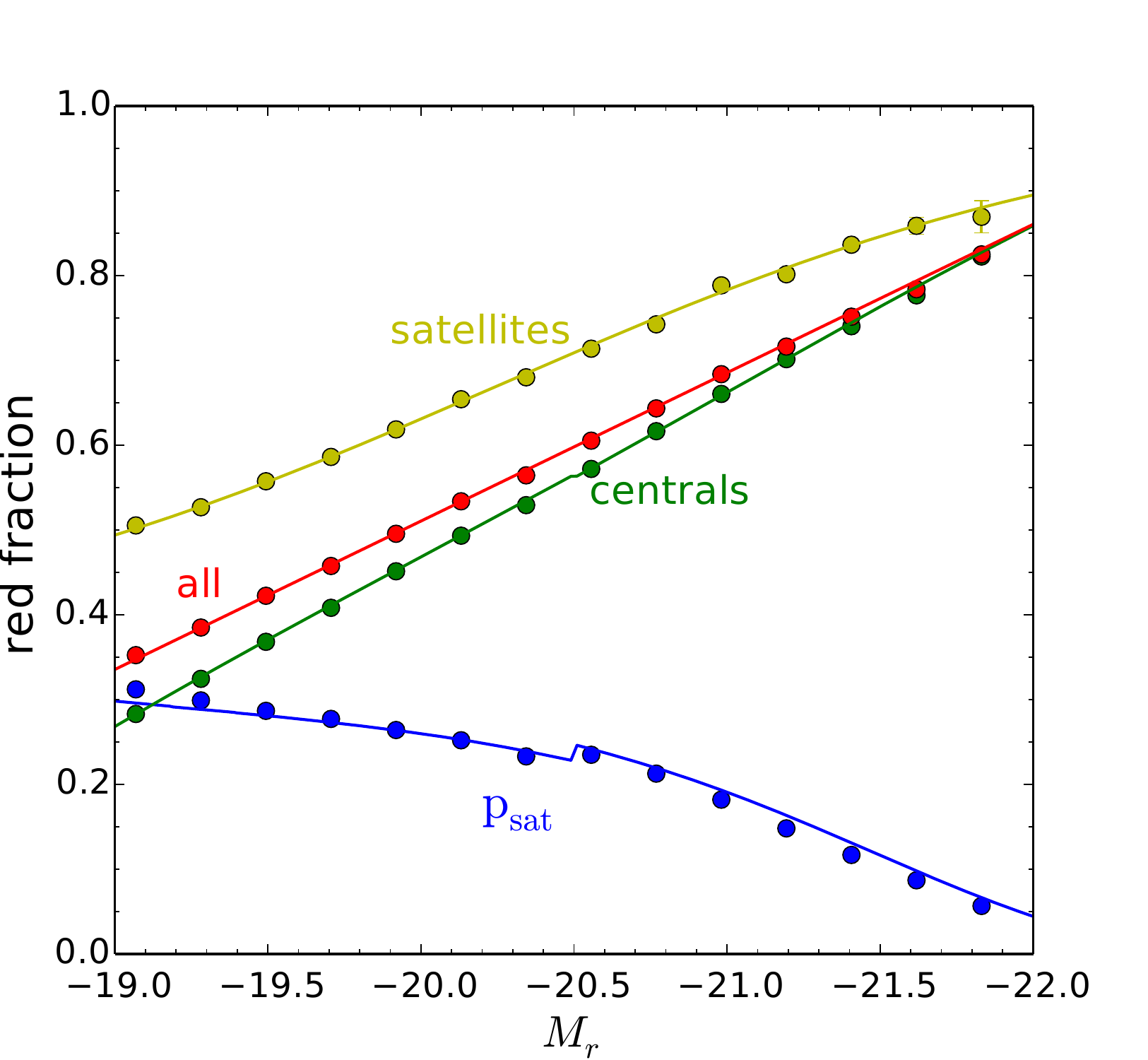}
\caption{Red fractions as a function of luminosity: The red, green and yellow lines respectively show the red fractions for all galaxies ($p_{\rm red}$), only centrals ($p_{{\rm red}|{\rm cen}}$) and only satellites ($p_{{\rm red}|{\rm sat}}$), as marked by the labels. For comparison, the satellite fraction $p_{\rm sat}$ \citep[equation~\ref{pbarsat} using the HOD calibrated by][]{zehavi+2011} is shown as the blue line. The filled symbols on top of all lines show the results of averaging over $10$ realisations of mocks given by P15. The error bars on the mock results show the standard deviation around the mean value. As expected, our analytical fractions agree with those in the mocks.}
\label{fig:redfrcs}
\end{figure}

The model is then completely specified if we know two out of $p({\rm red}|L)$, $p({\rm red}|{\rm sat},L)$ and $p({\rm red}|{\rm cen},L)$. Following P15, the first is fixed by fitting double Gaussians to the SDSS colour distribution at fixed luminosity and the second by matching the red fraction of satellites in the group catalog of \citet[][hereafter, Y07]{yang+07}:
\begin{align}
p({\rm red}|M_r) &= 0.423 - 0.175\left(M_r+19.5\right)\,, \notag\\
p({\rm red}|{\rm sat},M_r) &= 1.0 - 0.33\left[1+\tanh\left(\frac{M_r+20.25}{2.1}\right)\right]\,,
\label{dbl-Gauss-fits}
\end{align}
which then fix $p({\rm red}|{\rm cen},L)$ using \eqn{cen-sat-split-avg}. As a check, Figure~\ref{fig:redfrcs} compares the analytical forms above (together with the HOD calibrated by Z11), with measurements of corresponding quantities in the mock catalogs of P15 which used these expressions in assigning galaxy properties. 

Having fixed the various red fractions, the HOD split by galaxy colour can be described using
\begin{align}
\frcen(L|m) &\equiv p({\rm red}|{\rm cen},L) \fcen(L|m) \,,\notag\\
\cNrs(L|m) &\equiv p({\rm red}|{\rm sat},L) \cNs(L|m)\,,\notag\\
\Nrs(L|m) &\equiv \cNrs(L|m) / \fcen(>L|m), \notag \\
&= p({\rm red}|{\rm sat},L) \Ns(L|m)\,,
\label{colHOD-diff}
\end{align}
so that $\frcen(L|m)\der L$ is the differential fraction of $m$-haloes with red $L$-centrals and $\Nrs(L|m)\der L$ is the mean differential number of red $L$-satellites in $m$-haloes with centrals (of any colour) brighter than $L$. The corresponding cumulative HODs split by colour are
\begin{align}
\frcen(>L|m) &\equiv \int_{L}^{\infty}\der L^\prime\,\frcen(L^\prime|m)\,,\notag\\
\cNrs(>L|m) &\equiv \int_{L}^{\infty}\der L^\prime\,\cNrs(L^\prime|m)\,,\notag\\
\Nrs(>L|m) &\equiv \cNrs(>L|m) / \fcen(>L|m)\,.
\label{colHOD-cum}
\end{align}
Blue galaxies are described by similar functions, using $p({\rm blue}) = 1 - p({\rm red})$. For brevity, in the following, we will use the notation $\fcen(m)$ to mean, respectively, $\fcen(>L|m)$ and $\fcen(L|m)$ for samples thresholded and binned in luminosity, with similar abbreviations for \Ns, \cNs\ as well as all these quantities for the colour selected samples. 

\subsection{HOD with galactic conformity}
\label{sec:global:subsec:conf}
\noindent
We can now introduce conformity between the central and satellite colours in a group using the model of P15, which correlates the red fraction of a given galaxy type with the parent halo concentration. We briefly recapitulate this model and then describe the HOD including conformity.

Motivated by the fact that the distribution of halo concentration $c$ at fixed halo mass is approximately Lognormal with mean $\bar c(m)\equiv\,{\rm e}^{\avg{\ln c|m}}$ and logarithmic scatter $\sigma_{\ln c}$ \citep{st04,wechsler+06,dk15}, it is convenient to describe the following variable
\be
s\equiv\,\ln(c/\bar{c}) / \sigma_{\ln c}\,,
\label{eq:sdef}
\ee
whose distribution is approximately a standard Gaussian (zero mean and unit variance), \emph{independent} of halo mass\footnote{The relation $\bar c(m)$ has been calibrated by several authors. We will follow P15 and use the calibration of \citet{ludlow+14}, which is well described by $\bar c(m,z)=9.0\,\nu(m,z)^{-0.4}$ over the halo masses of interest \citep{paranjape14}, where $\nu(m,z)=\delta_{\rm c}(z)/\sigma(m)$ is the usual `peak height' defined using the spherical collapse density threshold $\delta_{\rm c}(z)$ and the linear theory r.m.s. $\sigma(m)$ of density fluctuations smoothed at mass scale $m$. The slightly different normalisation used by P15 was a result of an error in the code used by those authors, which we have fixed here and which does not affect any of the findings of P15.
The scatter $\sigma_{\ln c}\simeq0.14\ln(10)$ has been reported to be approximately independent of halo mass \citep{wechsler+06}. We use these expressions for the mean and scatter of concentration in all the results in the present work.}. 

P15 then introduce a concentration-dependence of the red fraction of satellites/centrals at fixed luminosity and halo mass by making galaxies in high (low) concentration haloes preferentially red (blue). The strength of this preference is determined using a constant parameter $\rho$ that lies between zero and unity, and which P15 interpreted as a ``group quenching efficiency'', in analogy with similar quenching efficiency parameters used in the literature previously \citep{vdb+08,peng+10,knobel+15}. In practice, one uses
\begin{align}
p({\rm red |sat/cen},s) &= (1-\rho)\, p({\rm red|sat/cen}) \notag\\
&\ph{(1-\rho)\,p()}
+ \rho\,\Theta(s-s_{\rm r,sat/cen})\,,
\label{predgivens-satcen}
\end{align}
where $s_{\rm r,sat/cen}$ gives the dividing line between high and low concentrations in each case as
\be
s_{\rm r,sat/cen} = \sqrt{2}\,{\rm erfc}^{-1} \left[ 2\,p({\rm red|sat/cen}) \right]\,,
\label{sred}
\ee
which follows from taking the distribution $p(s|m)=p(s)$ of the log-concentration to be Gaussian with zero mean and unit variance and demanding $\avg{p({\rm red|sat/cen,s)}}= p({\rm red|sat/cen})$. The extremes of the model are given by $\rho = 0$, the uncorrelated case when the red fraction does not depend on parent halo concentration, and $\rho = 1$ which corresponds to complete correlation where all galaxies in high (low)-concentration haloes are red (blue).

We can extend the previous analysis of colour-selected HODs to deal with $(m,s)$-haloes instead of $m$-haloes, thereby keeping track of halo mass and concentration simultaneously. In this case, we have
\begin{align}
\frcen(L|m,s) &\equiv p({\rm red|cen},L,s) \fcen(L|m) \,,\notag\\
\cNrs(L|m,s) &\equiv p({\rm red|sat},L,s) \cNs(L|m)\,,\notag\\
\Nrs(L|m,s) &\equiv \cNrs(L|m,s) / \fcen(>L|m) \notag \\
&= p({\rm red|sat},L,s) \Ns(L|m)\,,
\label{confcolHOD-diff}
\end{align}
where $\frcen(L|m,s)$ is the fraction of $(m,s)$-haloes with a red $L$-central and $\Nrs(L|m,s)$ is the mean number of red $L$-satellites in an $(m,s)$-halo that has a central brighter than $L$. The corresponding thresholded quantities are given by integrals over luminosity as before
\begin{align}
\frcen(>L|m,s) &\equiv \int_{L}^{\infty}\der L^\prime\,\frcen(L^\prime|m,s)\,,\notag\\
\cNrs(>L|m,s) &\equiv \int_{L}^{\infty}\der L^\prime\,\cNrs(L^\prime|m,s)\,,\notag\\
\Nrs(>L|m,s) &\equiv \cNrs(>L|m,s) / \fcen(>L|m)\,,
\label{confcolHOD-cum}
\end{align}
so that $\frcen(>L|m,s)$ is the fraction of $(m,s)$-haloes with a red central brighter than $L$ and $\Nrs(>L|m,s)$ is the mean number of red satellites brighter than $L$ in an $(m,s)$-halo with a central brighter than $L$. It is straightforward to show that averaging the quantities in \eqns{confcolHOD-diff} and~\eqref{confcolHOD-cum} over $s$ leads to the corresponding quantities in \eqns{colHOD-diff} and~\eqref{colHOD-cum}, respectively.

\subsection{1-halo conformity}
\label{sec:conf:subsec:1hconf}
\noindent
We can use the formalism above to calculate the red fraction of satellites with red or blue centrals; any difference in these two red fractions in groups of similar halo mass is a diagnostic of 1-halo conformity \citep{weinmann+06a,hearin+15a}. 
For example, we want to evaluate the probability $p(\textrm{sat is red}|L_{\rm s},\textrm{cen is red})$ that a satellite is red given that its central is red, for satellite luminosity $L_{\rm s}$, where it is understood that we are talking about galaxies in the same halo. We can calculate this as
\begin{align}
&p(\textrm{sat is red}|L_{\rm s},\textrm{cen is red}) \notag \\
&= \frac{\textrm{no. density of red sats w/ red cens}}{\textrm{no. density of sats w/ red cens}}\notag\\
&= \frac{\int\der m\,n(m)\int\der s\, p(s|m)\,\frcen(>L_{\rm s}|m,s)\,\Nrs(L_{\rm s}|m,s)}{\int\der m\,n(m)\int\der s\, p(s|m)\,\frcen(>L_{\rm s}|m,s)\,\Ns(L_{\rm s}|m)}\,,
\label{p(satred|cenred)}
\end{align}
Equation~\eqref{p(satred|cenred)} says that the red fraction of satellites with red centrals is simply the average red fraction of satellites, weighted by the number density of red centrals in $(m,s)$-haloes. Conformity arises due to the fact that the satellite red fraction and red central density both depend on the common value of the halo concentration $s$. Equation~\eqref{p(satred|cenred)} can be considerably simplified using our chosen model for conformity. A straightforward calculation shows that we can write
\begin{align}
&p(\textrm{sat is red}|L_{\rm s},\textrm{cen is red}) \notag\\
&= \bar p_{\rm r|sat} + \rho^2(1-\bar p_{\rm r|sat})\, \notag\\
&\ph{-\rho^2}
+ \rho^2\int\frac{\der m\,n(m)}{\bar n_{\rm sat,rc}}\,\Ns(L_{\rm s}|m)\bigg\{\bar p_{\rm r|sat}\fcen(>L_{\rm eq}|m)\notag\\
&\ph{1-\rho^2\int\frac{\der m\,n(m)}{\bar n_{\rm sat,rc}}\,\Ns(L_{\rm s}|m)[]}
-\frcen(>L_{\rm eq}|m)\bigg\}\,,
\label{p(satred|cenred)-simple}
\end{align}
where $\bar n_{\rm sat,rc}=\int\der m\,n(m)\,\Ns(L_{\rm s}|m)\,\frcen(>L_{\rm s}|m)$ is the number density of satellites with red centrals, $\bar p_{\rm r|sat} = p({\rm red}|{\rm sat},L_{\rm s})$, and we defined the function $L_{\rm eq}(L_{\rm s})$ which gives the luminosity at which the red fraction of centrals equals $\bar p_{\rm r|sat}$: 
\be
p({\rm red}|{\rm cen},L_{\rm eq}(L_{\rm s})) \equiv p({\rm red}|{\rm sat},L_{\rm s})\,.
\label{Leq-def}
\ee
A similar expression holds for the red fraction of satellites with blue centrals,
\begin{align}
&p(\textrm{sat is red}|L_{\rm s},\textrm{cen is blue}) \notag \\
&=\frac{\int\der m\,n(m)\int\der s\, p(s|m)\,\fbcen(>L_{\rm s}|m,s)\,\Nrs(L_{\rm s}|m,s)}{\int\der m\,n(m)\int\der s\, p(s|m)\,\fbcen(>L_{\rm s}|m,s)\,\Ns(L_{\rm s}|m)}\,,
\label{p(satred|cenblue)}
\end{align}
where $\fbcen(>L_{\rm s}|m,s) = \fcen(>L_{\rm s}|m) - \frcen(>L_{\rm s}|m,s)$. This can also be simplified to give
\begin{align}
&p(\textrm{sat is red}|L_{\rm s},\textrm{cen is blue}) \notag\\
&\ph{p(red)}
= p_{\rm rs|rc} - (\bar n_{\rm sat}/\bar n_{\rm sat,bc}) \left(p_{\rm rs|rc} - \bar p_{\rm r|sat}\right),
\label{p(satred|cenblue)-simple}
\end{align}
where $\bar n_{\rm sat,bc}=\int\der m\,n(m)\,\Ns(L_{\rm s}|m)\,\fbcen(>L_{\rm s}|m)$, and we set $p_{\rm rs|rc} = p(\textrm{sat is red}|L_{\rm s},\textrm{cen is red})$ as given by \eqn{p(satred|cenred)-simple}. 

A straightforward analysis shows that, for $\rho > 0$, at the faint end we must have 
\be
p(\textrm{sat is red}|L_{\rm s},\textrm{cen is red}) > p(\textrm{sat is red}|L_{\rm s},\textrm{cen is blue})\,,
\notag
\ee
a distinct signature of 1-halo conformity. (E.g., this is most easily seen upon setting $\rho=1$.) Setting $\rho=0$ on the other hand will lead to these fractions being equal to $p({\rm red}|{\rm sat},L_{\rm s})$ and hence to each other, \emph{provided} we assume the S09 model in which this concentration-averaged satellite red fraction does not depend explicitly on halo mass. If it does, then the mass averaging will lead to conformity-like effects even when $\rho=0$.

It is important to note that if one only allows $\fcen(L|m,s)$ and $\Ns(L|m,s)$ given $p({\rm red|sat},L)$ and $p({\rm red|cen},L)$ in equation~\eqref{confcolHOD-diff} and solves for equation~\eqref{p(satred|cenred)}, then the mass and $s$ independent factors like $p({\rm red|sat},L)$ will come out of the integrals and the rest in numerator and denominator cancel out.  In that case, $p(\textrm{sat is red}|L_{\rm s},\textrm{cen is red})=p({\rm r|sat})$, which implies that there will be no 1-halo conformity if one only allows $\fcen(L|m,s)$ and $\Ns(L|m,s)$.

\subsection{Galactic conformity and halo assembly bias}
\label{subsec:assemblybias}
\noindent
Halo assembly history is known to correlate with halo environment. This `assembly bias' manifests as a trend at fixed halo mass, in which low-mass older haloes tend to cluster more strongly at scales $\gtrsim 10$Mpc than younger haloes of similar mass, with the trend reversing for more massive haloes \citep{st04,gsw05,wechsler+06,jing+07,dalal+08,desjacques08a,hahn+09,fm10,fw10}. Further, halo concentration correlates positively with halo age~\citep{navarro+97,wechsler+02} and the assembly bias has been seen to extend to samples split by halo concentration, with more concentrated low-mass haloes clustering more strongly than less concentrated ones. As mentioned previously, this was the motivation for P15 to use halo concentration as the variable (in addition to halo mass) that determines galaxy colour.  

For the purposes of our analytical model, we therefore require a description of the mass \emph{and} concentration dependence of the linear large scale bias, $b(m,s)$. A fitting function for the $s$-dependence of this quantity has been provided by \citet{wechsler+06}, who use the parametrisation $b(m,s)\,=\,b(m)\,b_{c}(s|m/m_\ast)$, where $m_\ast(z)$ is the characteristic mass scale for collapse at redshift $z$, satisfying the relation $\nu(m_\ast,z)=1$. We have found, however, that this fitting function (equation~\ref{eq:wechr-fit}) does not describe very well the halo assembly bias in the simulations used by P15. In Appendix~\ref{app:bias-comp}, we provide a simple (but approximate) fix for the discrepancy. \footnote{We have also tried the fitting function as given in \cite{PP16}. However, this also has problems in describing the trends of halo assembly bias in the simulations used by P15. The reason could be that this fit is suitable for very large scale assembly bias. In our case, we are talking about comparatively smaller scales, (3-20 $h^{-1}$ Mpc). At those scales, there might be other effects contributing, for example, scale dependent all halo as well as assembly bias. These effects are not captured by the fits given by \cite{PP16}.} This highlights the need, however, of obtaining accurate calibrations of halo assembly bias for use in Halo Model analyses. The concentration-averaged version of linear bias $b(m)$, on the other hand, has been well studied in the literature, and we will use the fitting function provided by \citet{Tinker10}.

A model of galactic conformity driven by halo concentration would naturally manifest assembly bias effects in the spatial clustering of colour-split galaxy samples. While it is reasonable to expect that star formation activity  correlates with halo assembly, observationally it has been very challenging to establish the existence of \emph{galaxy} assembly bias at the faint end \citep[see, e.g.,][]{wwdy13,lin+16,twcm16}. One cannot, therefore, \emph{a priori} exclude the possibility that the 1-halo conformity reported by \citet{weinmann+06a} is due to some group-scale physical process that affects the centrals and satellites of a group, but is \emph{uncorrelated} with the large scale environment. To this end, P15 also explored a version of their model in which 1-halo conformity was left intact, but all assembly-bias effects were erased by scrambling halo concentrations among haloes of fixed mass. This `\emph{no-2h}' model was, in fact, able to qualitatively explain the conformity trends seen by \citet{kauffmann+13} at $\sim4$Mpc separations as potentially arising purely from the 1-halo effects of the most massive haloes in the volume. In our analytical framework, this \emph{no-2h} model can be reproduced by simply replacing $b(m,s)\to b(m)$ in all of the correlation functions described in the next section.

\section{Correlation functions}
\label{sec:corrfunc}
\noindent
Our main focus in this work is on analytically understanding the potential impact of galactic conformity, with or without assembly bias, in the spatial clustering of galaxies. Ideally, one would work at fixed halo mass so as to remove any mass-driven trends in red fractions and clustering; these tend to be strong and can easily confuse any analysis that looks for trends driven by variables \emph{other} than halo mass. This is, of course, straightforward to do in an analytical calculation or in mock catalogs. Unfortunately, controlling for halo mass using group catalogs based on  \emph{observations} is subject to rather strong systematic uncertainties \citep[see, e.g.,][]{campbell+15}. We will therefore focus on traditional 2-point statistics, which typically split galaxies into red and blue at fixed luminosity or stellar mass but \emph{do not} explicitly identify centrals and satellites on an object-by-object basis. We will then rely on our analytical formalism (backed by the P15 mock catalogs) to look for potential signatures of conformity in large scale clustering. Any such signatures, if robustly detectable, would be free from systematic uncertainty in identifying centrals/satellites or halo mass.

The basic statistic of interest is the 2-point correlation function $\xi(r)$ which satisfies
\be
\ngal^2 (1+\xi(r)) = \langle n_{\rm g} (\mathbf{x}) n_{\rm g}(\mathbf{x}+\mathbf{r})\rangle,
\ee
where $n_{\rm g}(\mathbf{x})$ is the galaxy number density at any point $\mathbf{x}$. The global (i.e., not colour-selected) galaxy number density of a luminosity thresholded/binned sample is given by
\be
\ngal = \int\der m\,n(m)\,\bigg[\fcen(m) + \cNs(m)\bigg]\,.
\label{ngal}
\ee
The galaxy 2-point correlation function $\xi(r)$ can be written as
\be
\xi(r) = \xi^{\rm (1h)}(r) + \xi^{\rm (2h)}(r),
\ee
and the $\xi^{\rm (1h)}(r)$ (1-halo) and $\xi^{\rm (2h)}(r)$ (2-halo) terms in spatial 3-d configuration/Fourier space can be calculated along the lines discussed by \citet{cs02}.

When comparing with SDSS measurements, we will also use the projected correlation function $w_{\rm p}(r_{\rm p})$ \citep{dp83} 
\begin{align}
w_{\rm p}(r_{\rm p}) &= \int_{-\infty}^\infty\der \pi\,\xi(r_{\rm p},\pi) = 2 \int_{r_{\rm p}}^{\infty}\frac{\der r\,r\,\xi(r)}{\sqrt{r^2 - r_{\rm p}^2}}\,,
\label{eq:wp_rp}
\end{align}
where $r_{\rm p}$ ($\pi$) is the separation between two galaxies perpendicular (parallel) to the line of sight. 

Here, we first briefly recapitulate the framework for calculating the all-galaxy correlation function, explicitly accounting for the scatter in the halo concentrations at fixed mass. This will set the stage for discussing the colour-split correlation functions, with and without conformity.

\subsection{All-galaxy correlation function}
\label{subsec:allgalcorrfunc}
\noindent
For Poisson distributed satellite counts, the all-galaxy 1-halo and 2-halo terms can be written as
\begin{align}
\xi^{\rm (1h)}(r) &= \frac{1}{\ngal^2} \int\frac{\der m\,n(m)}{\fcen(>L|m)} \int\der s\,p(s|m) \notag\\
&\ph{\frac{1}{\ngal^2}}
\bigg[2\,\fcen(m)\,\cNs(m) \frac{\rho(r|m,s)}{m} \notag\\
&\ph{\frac{1}{\ngal^2}2\,\fcen(m)}
+ \cNs(m)^2 \, \frac{\lambda(r|m,s)}{m^2} \bigg]\,,
\label{xi1h}\\
P^{\rm (2h)}(k) &= P_{\rm m}(k) \bbar(k)^2\,;
\label{P2h}\\
\bbar(k) &\equiv \frac{1}{\ngal} \int\der m\, n(m)\int\der s\,p(s|m)\,b(m,s) \notag \\
&\ph{\frac{1}{\ngal} \int\der m} 
\bigg\{\fcen(m)+\cNs(m)u(k|m,s)\bigg\}\,, 
\label{bk}
\end{align}
where $\rho(r|m,s)$ is a spherically symmetric, truncated halo profile calculated using the standardized concentration $s$ (equation~\ref{eq:sdef}), with $u(k|m,s)$ the Fourier transform and $\lambda(r|m,s)$ the convolution of this profile with itself\footnote{Throughout, we use an NFW profile \citep*{navarro+97} for $\rho(r|m,s)$.}. We have also assumed that the halo auto correlation can be approximated as $\xi_{\rm hh}(r|m,s,m',s')\approx b(m,s)b(m',s')\xi_{\rm m}(r)$, where $b(m,s)$ is the linear halo bias (the $s$-dependence of which is due to halo assembly bias, see section~\ref{subsec:assemblybias}) and $\xi_{\rm m}(r)$ and $P_{\rm m}(k)$ are the correlation function and power spectrum, respectively, of dark matter. Recall that any power spectrum $P(k)$ is the Fourier transform of the corresponding correlation function $\xi(r) = \int\der k\,k^2 \,P(k)\,\sin(kr)/(kr)/(2\pi^2)$. 

The expression for the 2-halo correlation can be improved by including the effects of nonlinearities in the dark matter power spectrum, halo exclusion and the scale dependence of halo bias \citep{smith+03,tinker+05,takahashi+12,vdb+13}. While we use the nonlinear fitting function HALOFIT for $P_{\rm m}(k)$ \citep{smith+03} and account for the scale dependence of halo bias using the prescription of \citet{tinker+05}, in this work we ignore the effects of halo exclusion. We comment on the systematics introduced by this choice later.

It is worth pointing out that the existence of all galaxy/halo assembly bias might be uncorrelated with the conformity between central and satellite galaxies. As there is no dependence of HOD on $s$ in \eqns{xi1h}-\eqref{bk}, all galaxy assembly bias is simply a consequence of halo assembly bias.  However, this may not be true if central-satellite conformity is present. Thus, in general, one should allow both $\fcen(m,s)$ and $\Ns(m,s)$ regardless of presence of conformity. In that case, the above equations will become

\begin{align}
\xi^{\rm (1h)}(r) &= \frac{1}{\ngal^2} \int\der m\,n(m)\int \frac{\der s\,p(s|m)}{\fcen(>L|m,s)}  \notag\\
&\ph{\frac{1}{\ngal^2}}
\bigg[2\,\fcen(m,s)\,\cNs(m,s) \frac{\rho(r|m,s)}{m} \notag\\
&\ph{\frac{1}{\ngal^2}2\,\fcen(m)}
+ \cNs(m,s)^2 \, \frac{\lambda(r|m,s)}{m^2} \bigg]\,,
\label{xi1h-s}\\
P^{\rm (2h)}(k) &= P_{\rm m}(k) \bbar(k)^2\,;
\label{P2h-s}\\
\bbar(k) &\equiv \frac{1}{\ngal} \int\der m\, n(m)\int\der s\,p(s|m)\,b(m,s) \notag \\
&\ph{\frac{1}{\ngal} \int\der m} 
\bigg\{\fcen(m,s)+\cNs(m,s)u(k|m,s)\bigg\}\,.
\label{bk-s}
\end{align}
\noindent
\cite{as07} showed that the mock galaxy catalogues based on standard halo model description, constructed to reproduce the all galaxy clustering, exhibits approximately the same environmental-dependent clustering signals as seen in the SDSS. Thus, we take this as an approximation that all galaxy correlation functions matches well with that of SDSS and assume that $\fcen$ and $\Ns$ only depend on mass and luminosity.

\subsection{Colour-selected correlation functions}
\label{subsec:colourcorrfunc}
\noindent
The colour-selected correlation functions can be split into three categories, not all of which contain independent information. These are the auto-correlations of red and blue galaxies, respectively, and their cross-correlation. We first give the expressions for these without accounting for galactic conformity, which we discuss in the next section.

The number density of red galaxies, \nrgal, is given as
\begin{align}
\nrgal &= \int\der m\,n(m)\,\bigg[\frcen(m) + \cNrs(m)\bigg]\,,
\label{nrgal}
\end{align}
where $\frcen(m)$ and $\cNrs(m)$ were defined in \eqns{colHOD-diff} and~\eqref{colHOD-cum} , respectively. 

Similarly, one can write the auto-correlation of red galaxies as
\begin{align}
\xi_{\rm rr}^{\rm (1h)}(r) &= \frac{1}{\nrgal^2} \int\frac{\der m\,n(m)}{\fcen(>L|m)} \int\der s\,p(s|m) \notag\\
&\ph{\,}
\bigg[2\,\frcen(m)\,\cNrs(m) \frac{\rho(r|m,s)}{m} \notag\\
&\ph{\frac{1}{\ngal^2}2\,\fcen(m)}
+ \cNrs(m)^2 \, \frac{\lambda(r|m,s)}{m^2} \bigg]\,,
\label{xir1h}\\
P_{\rm rr}^{\rm (2h)}(k) &= P_{\rm m}(k) \brbar(k)^2\,,
\label{Pr2h}\\
\brbar(k) &\equiv \frac{1}{\nrgal} \int\der m\, n(m)\int\der s\,p(s|m)\,b(m,s) \notag \\
&\ph{\frac{1}{\nrgal}\int\der m n}
\bigg\{\frcen(m)+\cNrs(m)u(k|m,s)\bigg\}\,.
\label{brk}
\end{align}
The reason the global cumulative fraction $\fcen(>L|m)$ continues to appear in the colour selected 1-halo term is because satellites (colour-selected or not) are always counted in haloes occupied by a central at least as bright as the satellite in question, regardless of the central's colour.
The auto-correlation of blue galaxies can be written identically, with the replacement `r'$\to$`b' in \eqns{nrgal}-\eqref{brk}. 

The cross-correlation between red and blue galaxies can then be obtained from the following constraint relations (suppressing the $r$/$k$ dependence in the correlation functions/power spectra, respectively)
\begin{align}
\ngal^2\,\xi^{\rm (1h)} &= \nrgal^2\,\xi_{\rm rr}^{\rm (1h)} + \nbgal^2\,\xi_{\rm bb}^{\rm (1h)} + 2\,\nrgal\,\nbgal\,\xi_{\rm rb}^{\rm (1h)}\,,\notag\\
\ngal^2\,P^{\rm (2h)} &= \nrgal^2\,P_{\rm rr}^{\rm (2h)} + \nbgal^2\,P_{\rm bb}^{\rm (2h)} + 2\,\nrgal\,\nbgal\, P_{\rm rb}^{\rm (2h)}\,.
\label{autocrossconstraint}
\end{align}

\subsection{Correlation functions in the presence of galactic conformity}
\label{subsec:corrfuncconf}
\noindent
The formalism above generalises in a straightforward manner to include the effects of galactic conformity, with the simple replacement $\frcen(m)\to\frcen(m,s)$ and $\cNs(m)\to\cNs(m,s)$, with the $s$-dependence given by \eqns{confcolHOD-diff} and~\eqref{confcolHOD-cum} with a non-zero value of $\rho$. We have,
\begin{align}
\xi_{\rm rr}^{\rm (1h)}(r) &= \frac{1}{\nrgal^2} \int\frac{\der m\,n(m)}{\fcen(>L|m)} \int\der s\,p(s|m) \notag\\
&\ph{\frac{1}{\nrgal}}
\bigg[2\,\frcen(m,s)\,\cNrs(m,s) \frac{\rho(r|m,s)}{m} \notag \\
&\ph{\frac{1}{\nrgal}2\,\frcen(m,s)}
+ \cNrs(m,s)^2 \, \frac{\lambda(r|m,s)}{m^2} \bigg]\,,
\label{xir1h-conf}\\
P_{\rm rr}^{\rm (2h)}(k) &= P_{\rm m}(k) \brbar(k)^2\,,
\label{Pr2h-conf}\\
\brbar(k) &\equiv \frac{1}{\nrgal} \int\der m\, n(m)\int\der s\,p(s|m)\,b(m,s) \notag \\
&\ph{\frac{1}{\nrgal}\int\der}
\bigg\{\frcen(m,s)+\cNrs(m,s)u(k|m,s)\bigg\}\,,
\label{brk-conf}
\end{align}
where \nrgal\ is the same as defined in \eqn{nrgal} (since the concentration dependence averages out), but the $s$-dependence of the coloured HODs now combines nontrivially with that of the halo profile $\rho(r|m,s)$ as well as the assembly bias inherent in $b(m,s)$. These reduce to the expressions in \eqns{xir1h}-\eqref{brk} upon setting the conformity parameter $\rho=0$.

As before, the expressions for the auto-correlation of blue galaxies are identical to those of the red, with the replacement `r'$\to$`b', while the red-blue cross-correlation follows from \eqn{autocrossconstraint}; for completeness, we give the full expression here:
\begin{align}
\xi_{\rm rb}^{\rm (1h)}(r) &= \frac{1}{\nrgal\nbgal} \int\frac{\der m\,n(m)}{\fcen(>L|m)}\int\der s\,p(s|m)  \notag\\
&\ph{\,}
\bigg[\big\{\frcen(m,s)\,\cNbs(m,s) \notag\\
&\ph{\frac{1}{\nrgal}}
+  \fbcen(m,s)\,\cNrs(m,s)\big\} \frac{\rho(r|m,s)}{m} \notag\\
&\ph{\frac{1}{\nrgal\nbgal}}
+ \cNrs(m,s)\,\cNbs(m,s)  \frac{\lambda(r|m,s)}{m^2} \bigg]\,,
\label{xirb1h-conf}\\
P_{\rm rb}^{\rm (2h)}(k) &= P_{\rm m}(k) \, \brbar(k)\,\bbbar(k)\,,
\label{Prb2h-conf}
\end{align}
with $\brbar(k)$ and $\bbbar(k)$ now defined as in \eqn{brk-conf}.

\begin{figure}
\centering
\includegraphics[width=0.45\textwidth]{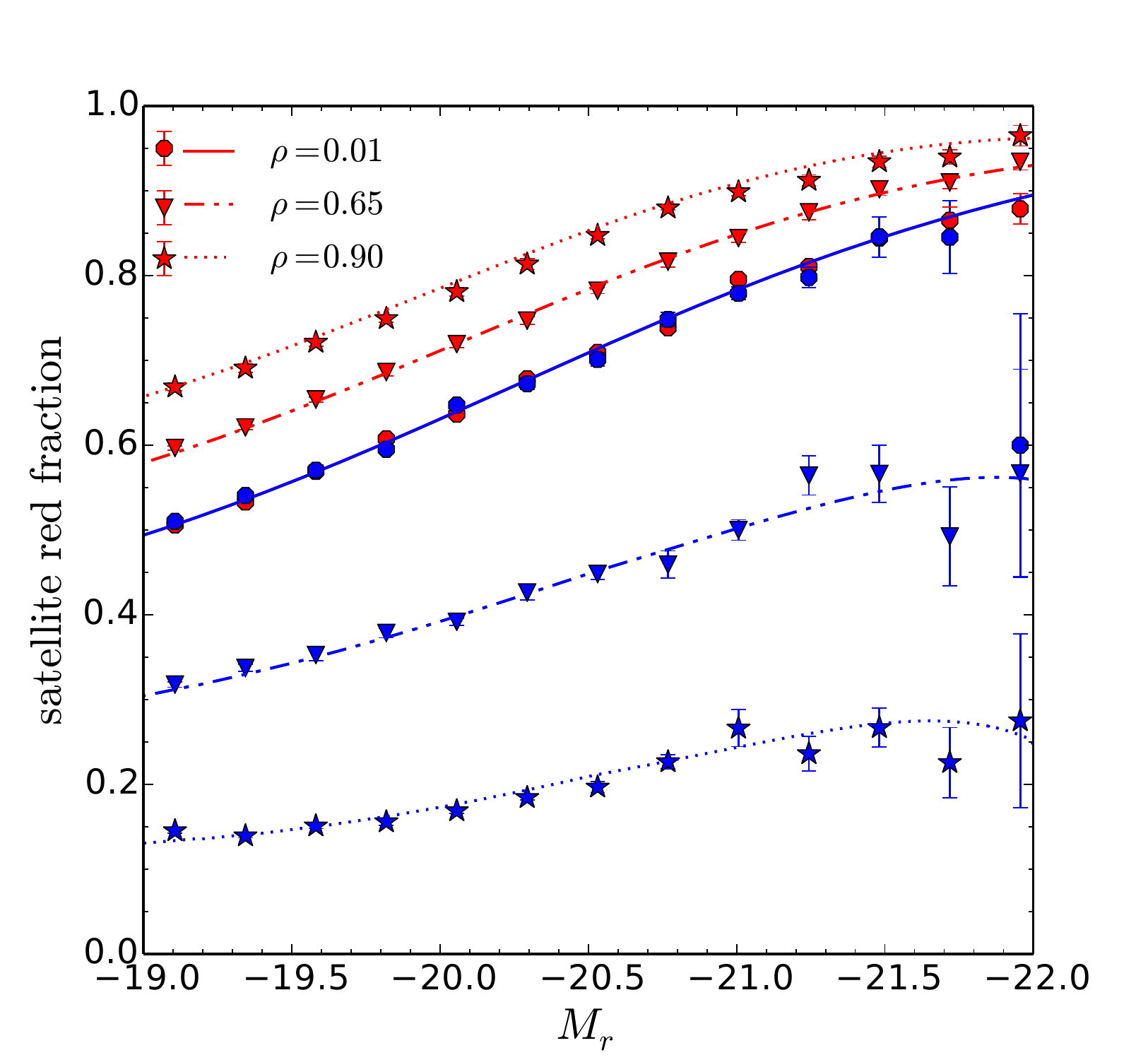}
\caption{Red fraction of satellites with red centrals (upper, red) and blue centrals (lower, blue) as a function of luminosity: Points with errors show measurements in the mock catalogs of \citet{pkhp15}, and smooth curves show the corresponding results of our analytical model. The measurements in the mocks were averaged over 10 independent realisations, with the error showing the standard deviation around the mean. Circles/solid lines show results for `no conformity' with $\rho=0.01$, triangles/dot-dashed lines are for `medium' conformity with $\rho=0.65$, while stars/dotted lines show results for `strong' conformity with $\rho=0.9$. See text for a discussion.}
\label{fig:1hc}
\end{figure}

\begin{figure*}
\centering
\includegraphics[width=0.75\textwidth]{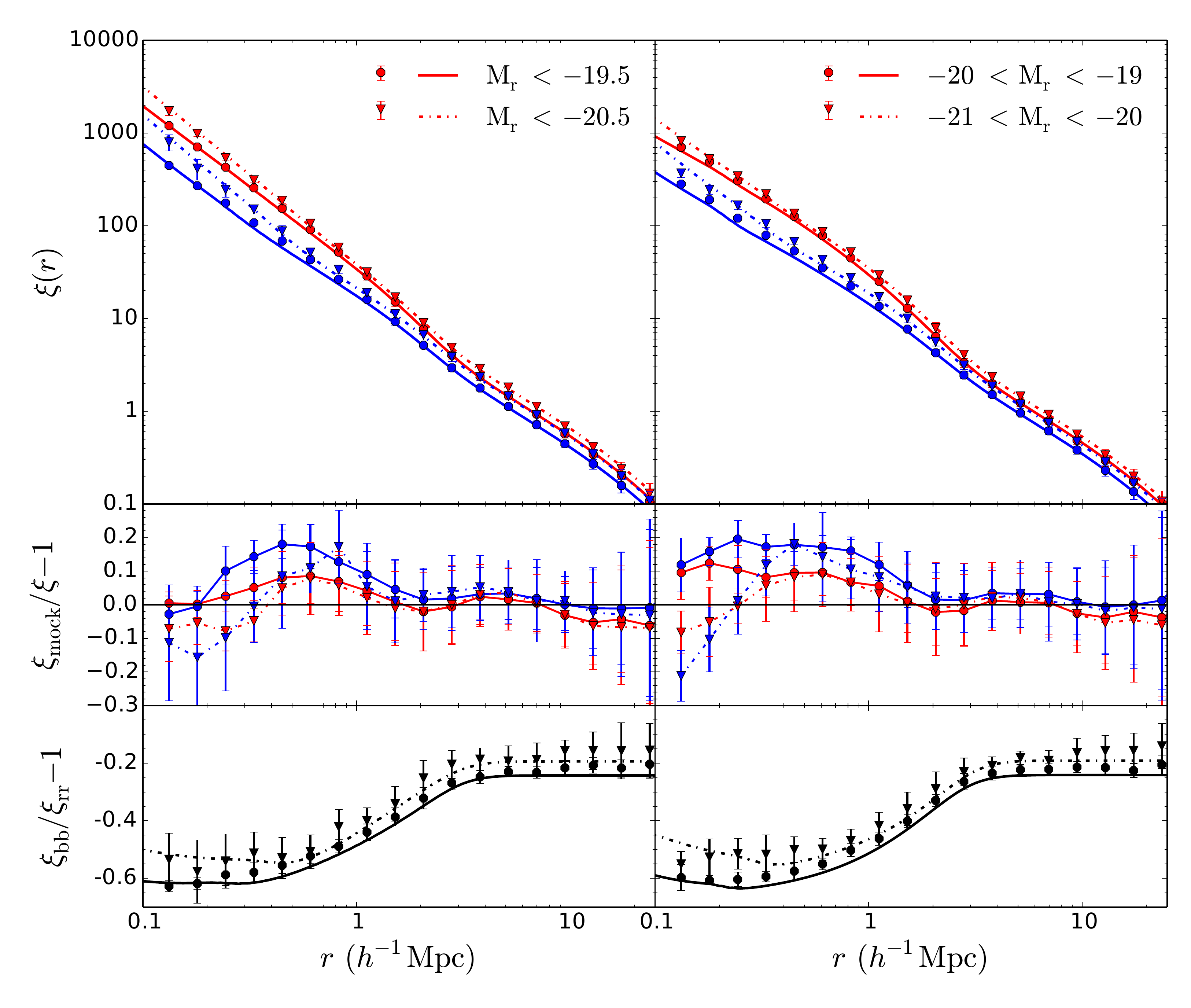}
\caption{This figure compares the analytical correlation function obtained from our method to the one obtained from mocks for red and blue galaxies (stronger and weaker clustering, respectively). We have used $\rho\,=\,0.01$ mocks for this comparison. The \emph{left} and \emph{right} panels correspond to the correlation functions for the samples thresholded and binned in luminosity, respectively. The lines represent the analytical results and the markers show the results from averaging measurements in 10 independent mocks. The different line-styles and markers represent the different luminosity threshold/bins as indicated. The \emph{top row} shows individual correlation functions, while the \emph{middle row} shows the residuals between corresponding analytical and numerical results. The \emph{bottom row} shows the relative clustering of blue and red galaxies: $\xi_{\rm bb}(r)/\xi_{\rm rr}(r)-1$. }
\label{fig:corrfn-scat}
\end{figure*}

\subsection{Correlation functions with alternative models}
In general, our HOD model can be even further extended by including other additional effects such as the dependence of $\cNrs$ and $\cNbs$ on the colour of the central galaxy. This type of dependency gives rise to an extra correlation between the colour of the central galaxy and the colour of the satellite galaxy. By taking an additional dependence on the colour of central galaxy, there will be a degeneracy between the conformity effects coming from concentration and central galaxy colour dependence. As the galactic conformity effect itself is very small, adding an additional dependence will make it difficult to distinguish between the individual contributions coming from both effects. Hence, we choose not to consider such extra correlations in this work and only consider model given in P15.

The other alternative model which can produce the 1-halo conformity like effects, can be obtained by changing the mean NFW concentration parameter. Till now, we have used the same spatial profile for the distribution of satellites for both red and blue galaxies. However, our model also allows the mean NFW concentration parameter for red galaxies to be different than blue ones.  While the choice of an NFW profile is well motivated \citep{berlind+03,zheng04}, it is known that satellites in the interiors of haloes tend to be redder than in the outskirts \citep{ms77,bo78,dkcw99,hswk09}. In principle, one could set up a machinery that allows for different spatial profiles for red and blue satellites \citep{scranton02}. In section \ref{sec:appl}, we also explore this possibility by considering a simpler model by allowing different mean concentration parameters for the two populations, while retaining the NFW form for their profiles (see also Z11). It is important to point out that the changes in the mean concentration parameters for red and blue galaxies by keeping other quantities fixed in the expressions of colour-selected correlation functions will only affect small scales \citep{sheth+01} and hence will not lead to 2-halo scale effects.

\section{Validation against mock catalogs}
\label{sec:results}
\noindent
The analytical expressions of the previous section can be validated by comparing with measurements of these observables in the mock catalogs described by P15. Briefly, these mocks are based on 10 realisations of an $N$-body simulation in a cubic periodic box of size $\left(200\Mpch\right)^3$, with a particle mass of $m_{\rm part}=4.1\times10^9\Mh$ and using haloes with $m\geq20m_{\rm part}$. Central galaxies are placed at the centers of mass of haloes and satellites are distributed around these using NFW profiles with Lognormal concentrations (see section~\ref{sec:global:subsec:conf}). The mocks are luminosity-complete for $M_r < -19.0$, with central and satellite luminosity functions constrained to satisfy the Z11 HOD. Galaxy colours are prescribed using the S09 algorithm, with red fractions as described in section~\ref{sec:global:subsec:redfrac}. To introduce the galactic conformity, the galaxy colours are correlated with the halo concentrations. Ideally, one would use the concentrations actually measured in the simulation. However, these concentrations are only approximately Lognormal while the conformity model of P15 assumes that they are exactly Lognormal. Thus, using the simulated concentrations directly, in general, will not preserve the central and satellite red fractions. It is important to note, however, that we are not really interested in the actual shape of the distribution of concentrations, but only their ranking in haloes of fixed mass. Thus, 1-halo conformity is induced by numerically implementing \eqn{predgivens-satcen}, and 2-halo conformity by rank ordering the Lognormal concentrations by the measured halo concentrations in bins of halo mass, prior to assigning satellite positions and galaxy colours. We refer the reader to P15 for more details. We start with a comparison of 1-halo conformity and then move on to measurements of spatial clustering.

\subsection{1-halo conformity}
\label{sec:results:subsec:1hconf}
\noindent
Figure~\ref{fig:1hc} compares the red fraction of satellites with red centrals with that of satellites with blue centrals, as functions of satellite luminosity for different values of $\rho$. The points with error bars show the measurements in the P15 mocks while the smooth curves show \eqns{p(satred|cenred)-simple} and~\eqref{p(satred|cenblue)-simple}. P15 discussed a similar plot as a function of stellar mass. We see that the difference between these two red fractions essentially vanishes for $\rho=0.01$ and is very large for $\rho=0.9$, with satellites being preferentially red (blue) if their central is red (blue) in the latter case. 
Throughout, we will use the values $\rho=0.01$ and $0.9$ to represent `no conformity' and `strong' conformity, respectively, and the value $\rho=0.65$ for `medium' conformity. This last value was shown by P15 to give a good description of 1-halo conformity in the group catalog of Y07. We can clearly see from the Figure that, as expected, our analytical model correctly tracks the behaviour of the conformity signal as $\rho$ is varied.

\begin{figure*}
\centering
\includegraphics[width=0.65\textwidth]{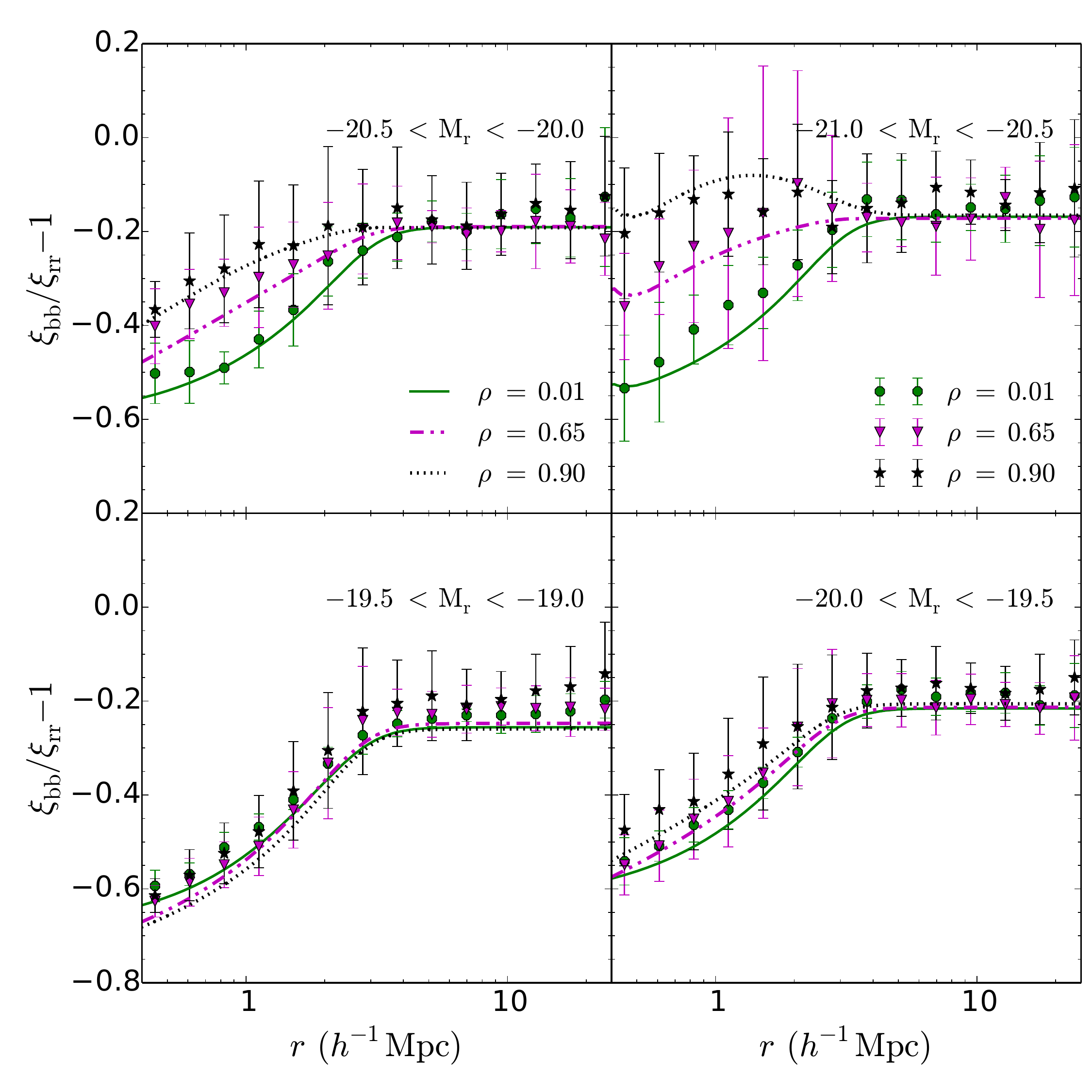}
\caption{The effect of introducing 1-halo conformity in the correlation function: This figure compares the excess of blue galaxy correlation function over that of the red for $\rho\,=\,0.01,0.65$ and $0.9$ represented by the solid green, dot-dashed magenta and dotted black lines. The different symbols are the mean of the corresponding results from 10 independent mocks with standard deviation as the error bars on them. The different panels correspond to the different luminosity bins as indicated.}
\label{fig:corr-noconf-thresh}
\end{figure*}

\begin{figure*}
\centering
\includegraphics[width=0.65\textwidth]{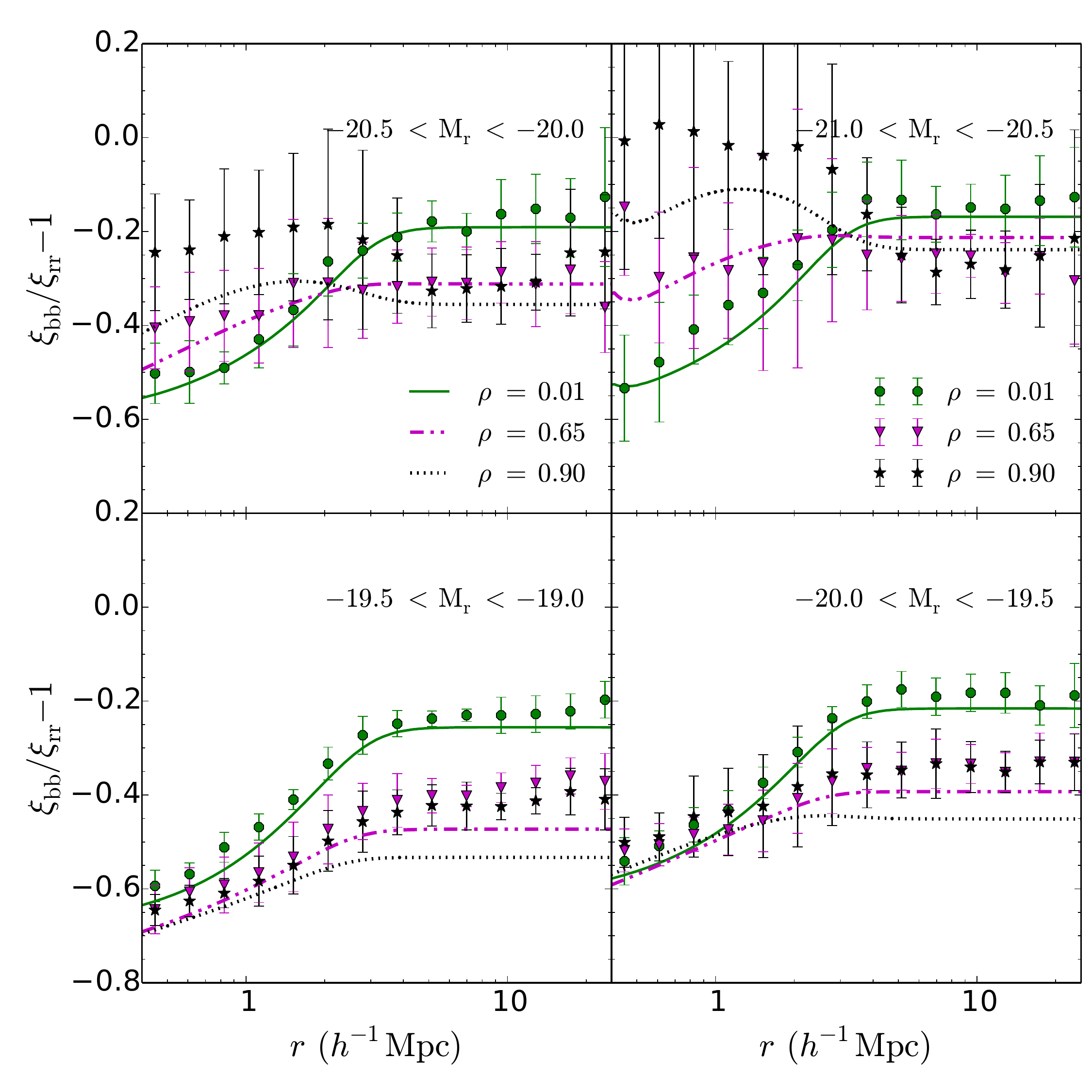}
\caption{The effect of introducing 2-halo conformity in the correlation function: Similar to Figure~\ref{fig:corr-noconf-thresh} but with 2-halo conformity switched on. See text for a discussion.}
\label{fig:corr-conf-thresh}
\end{figure*}

\subsection{Spatial clustering}
\label{sec:results:subsec:clust}
\noindent
The analytical model described in section~\ref{sec:corrfunc} does not account for halo exclusion, and is known to fail at the $\sim10$-$20$ per cent level in describing spatial clustering of any galaxy sample at separations of the order of the typical virial radius of the host dark matter halo \citep[see][for a comprehensive analysis]{vdb+13}. Since our analysis below will heavily rely on comparing the clustering of red and blue galaxies, whose host halo masses can be substantially different at fixed luminosity, we start by asking how well our model does in the \emph{absence} of conformity.

Figure~\ref{fig:corrfn-scat} compares $\xi(r)$ for red and blue galaxies (colour coded) in samples defined by luminosity thresholds (left panels) and bins (right panels). The points with errors in the top panels show measurements in the mocks using the estimator from \cite{ls93}, while the smooth curves show our analytical results using \eqns{nrgal}-\eqref{brk}. 
We see the well-known trend that red galaxies cluster more strongly than blue galaxies of the same luminosity at all separations. In the analytical model, this can be understood by studying the relative fractions of red and blue galaxies that are satellites. E.g., at the faint end $-19.5 < M_r < -19$, from Figure~\ref{fig:redfrcs} we have $p({\rm sat}|{\rm red}) \sim 0.43$ and $p({\rm sat}|{\rm blue}) \sim 0.23$, i.e., $\sim43\%$ of red galaxies are satellites, while only $\sim23\%$ of blue galaxies are satellites. This means that the population of faint blue galaxies is dominated by centrals in low-mass haloes, while that of faint red galaxies has a substantial number of satellites in higher mass haloes. Consequently, the large scale bias of the blue population is smaller than that of the red. At small enough scales, on the other hand, the correlation function is dominated by the satellite-satellite term in \eqn{xir1h} which is approximately proportional to $p({\rm sat}|{\rm red/blue})^2$, so that the fact that $p({\rm sat}|{\rm red}) > p({\rm sat}|{\rm blue})$ directly translates into stronger clustering for the red galaxies.

The middle panels of Figure~\ref{fig:corrfn-scat} show the residuals between the analytical $(\xi)$  and numerical $(\xi_{\rm mock})$ results; we see that these are at the $\sim10$-$20$ per cent level at intermediate scales, while dropping below $5$ per cent at small and large scales. These residuals are a direct estimate of the error incurred due to the approximations discussed in section~\ref{subsec:allgalcorrfunc}, in particular, the fact that we do not account for halo exclusion.

The bottom panels of Figure~\ref{fig:corrfn-scat} shows the relative difference between red and blue clustering, $\xi_{\rm bb}/\xi_{\rm rr} - 1$. This quantity has the dual advantage that sample variance effects in its numerical estimates largely cancel, while the systematic deviations in our analytical estimates are also somewhat suppressed. In the remainder of this section, we will therefore exclusively focus on this relative difference\footnote{This will, of course, hide any effect of conformity on the \emph{overall} strength of clustering, which is also of interest; we will therefore study the full correlation function later when comparing with SDSS measurements.}.
Given the quality of agreement seen in Figure~\ref{fig:corrfn-scat}, we expect our analytical model to work at the $\sim10$-$20$ per cent level at separations $0.2 \lesssim r/(\Mpch) \lesssim 20$.

\subsubsection{The `\emph{no-2h}' model}
\label{subsubsec:no2h}
\noindent
To investigate the effect of 1-halo conformity on the correlation functions, we use non-zero values of $\rho$ in \eqns{xir1h-conf}-\eqref{brk-conf} while setting $b(m,s) \to b(m)$ in \eqn{Pr2h-conf}.  
Figure~\ref{fig:corr-noconf-thresh} compares the resulting analytical predictions with measurements in the \emph{no-2h} mocks of P15 (see section~\ref{sec:global:subsec:conf} above), for galaxies in bins of luminosity. As before, we display results for $\rho=0.9,0.65,0.01$ corresponding to strong, medium and no conformity, respectively.

Overall, our analytical results correctly track the trends measured in the mocks. At fixed luminosity, we see that the magnitude of the relative difference between red and blue clustering decreases as $\rho$ increases, and picks up a nontrivial scale dependence. The trend with $\rho$ is quite mild at the faint end and becomes more dramatic for brighter samples. These trends are most easily understood upon comparing the extreme cases $\rho=0$ and $\rho=1$. In the former, no conformity case, at small enough separations, the red/blue correlation function scales like $\sim p({\rm sat}|{\rm red/blue})^2$, as discussed previously. In the strong conformity case with $\rho=1$, on the other hand, the population of satellites at fixed luminosity and halo mass gets completely segregated, with red (blue) satellites only occupying high (low) concentration haloes. This leads to increased small scale clustering for both red and blue satellites, but with a different increase in each case. To see this, note that, for $\rho\to1$, the satellite red/blue fraction becomes a 0/1 step function in halo concentration (equation~\ref{predgivens-satcen}). The corresponding small scale correlation functions now scale like $\sim p({\rm red/blue}|{\rm sat})p({\rm sat})^2/p({\rm red/blue})^2$, which are always \emph{larger} than $p({\rm sat}|{\rm red/blue})^2$, respectively, so that both clustering amplitudes increase. The change in the \emph{relative} clustering $\xi_{\rm bb} / \xi_{\rm rr}$ from $\rho=0$ to $\rho=1$, however, scales like $(\xi_{\rm bb} / \xi_{\rm rr})|_{\rho=1} / (\xi_{\rm bb} / \xi_{\rm rr})|_{\rho=0}\sim p({\rm red}|{\rm sat})/p({\rm blue}|{\rm sat})$, which is a mild difference at the faint end where the satellite blue fraction is close to 0.5, but becomes a large effect for brighter samples where the blue fraction falls towards zero (Figure~\ref{fig:redfrcs}).

\subsubsection{2-halo conformity}
\label{subsubsec:2hconf}
\noindent
We now switch on 2-halo conformity by allowing the galaxy colours to respond to a concentration-dependent halo bias $b(m,s)$. Since this assembly bias is qualitatively different at different halo masses, we expect a nontrivial trend in the relative large scale clustering of red and blue galaxies. This trend will be largely driven by the halo occupation statistics of central galaxies, which constitute the dominant population at any luminosity.

To see what to expect, recall that, at low halo mass, halo bias increases monotonically with halo concentration. At the faint end, for nonzero values of $\rho$, red galaxies will occupy preferentially more concentrated haloes than blue galaxies. Consequently, with increasing $\rho$, the large scale bias of red (blue) galaxies will be larger (smaller) than the corresponding bias when $\rho=0$; the magnitude of the relative clustering between faint red and blue galaxies will therefore increase with $\rho$. As the galaxy sample is made brighter, the relevant halo mass increases and approaches the inversion mass scale, where halo bias becomes nearly independent of concentration (see Appendix~\ref{app:bias-comp}). As a result, the trend with $\rho$ should become milder for brighter galaxies\footnote{Since our mocks do not have the volume to reliably sample very massive clusters, we do not expect to reliably probe the large mass regime where the assembly bias trend with halo concentration is inverted compared to that at smaller masses.}.

These trends are borne out by Figure~\ref{fig:corr-conf-thresh}, which is formatted identically to Figure~\ref{fig:corr-noconf-thresh} and compares measurements using the `default' mocks of P15 which included halo assembly bias (points with errors) with the analytical results of \eqns{xir1h-conf}-\eqref{brk-conf}; the $s$-dependence of halo bias in the latter is given by equation~\eqref{eq:wechr-fit} with the correction discussed in Appendix~\ref{app:bias-comp}. While the analytical model qualitatively captures the measured trends, the level of agreement at large scales is only $\sim30\%$ for the faintest galaxies. As we discuss in Appendix~\ref{app:bias-comp}, this disagreement is likely to be dominated by systematic uncertainties in the analytical description of halo assembly bias. This highlights the need for accurate calibrations of assembly bias for studies such as this one.

\section{Projected clustering in SDSS}
\label{sec:appl}
\begin{figure}
\centering
\includegraphics[width=0.45\textwidth]{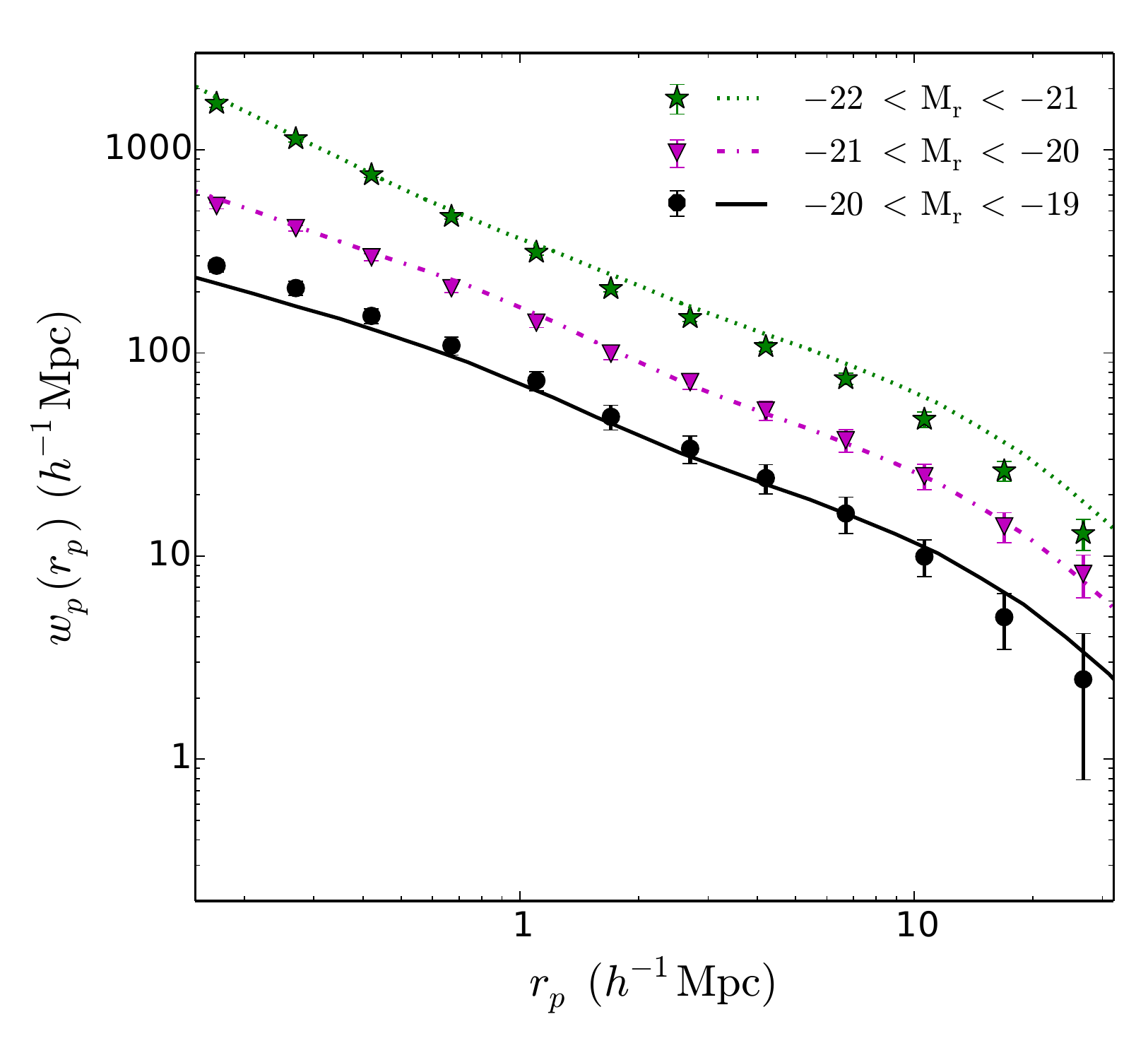}
\caption{All-galaxy projected correlation function: points with errors show the measurements from SDSS galaxies (Table 7 of Z11) and smooth lines show the results from our analytical model. For clarity, we have staggered the results for the two brighter bins upwards by 0.25dex.}
\label{fig:wprp-allgal}
\end{figure}

\noindent
We now turn to a comparison with the observed colour-dependent clustering of SDSS galaxies as quantified by the projected correlation function $w_{\rm p}(r_{\rm p})$ (equation~\ref{eq:wp_rp}) of red and blue galaxies in bins of luminosity (Z11)\footnote{We replace the upper limit of the first integral in \eqn{eq:wp_rp} with $\pi_{\rm max}=60 h^{-1}$Mpc, the same as used in Z11.}. Figure~\ref{fig:wprp-allgal} compares our analytical results for the all-galaxy projected correlation function in three luminosity bins with corresponding measurements from Z11, taken from their Table~$7$. We see that the model does a good job at all scales for faint and intermediate luminosities, but overpredicts the large scale clustering in the brightest bin. This is possibly due to an overestimated satellite fraction in the HOD calibration we are using. Fixing this discrepancy would require recalibrating the HOD, which is beyond the scope of this work. In the following, therefore, we will focus on the two fainter luminosity bins.

\begin{figure*}
\centering
\includegraphics[width=0.9\textwidth]{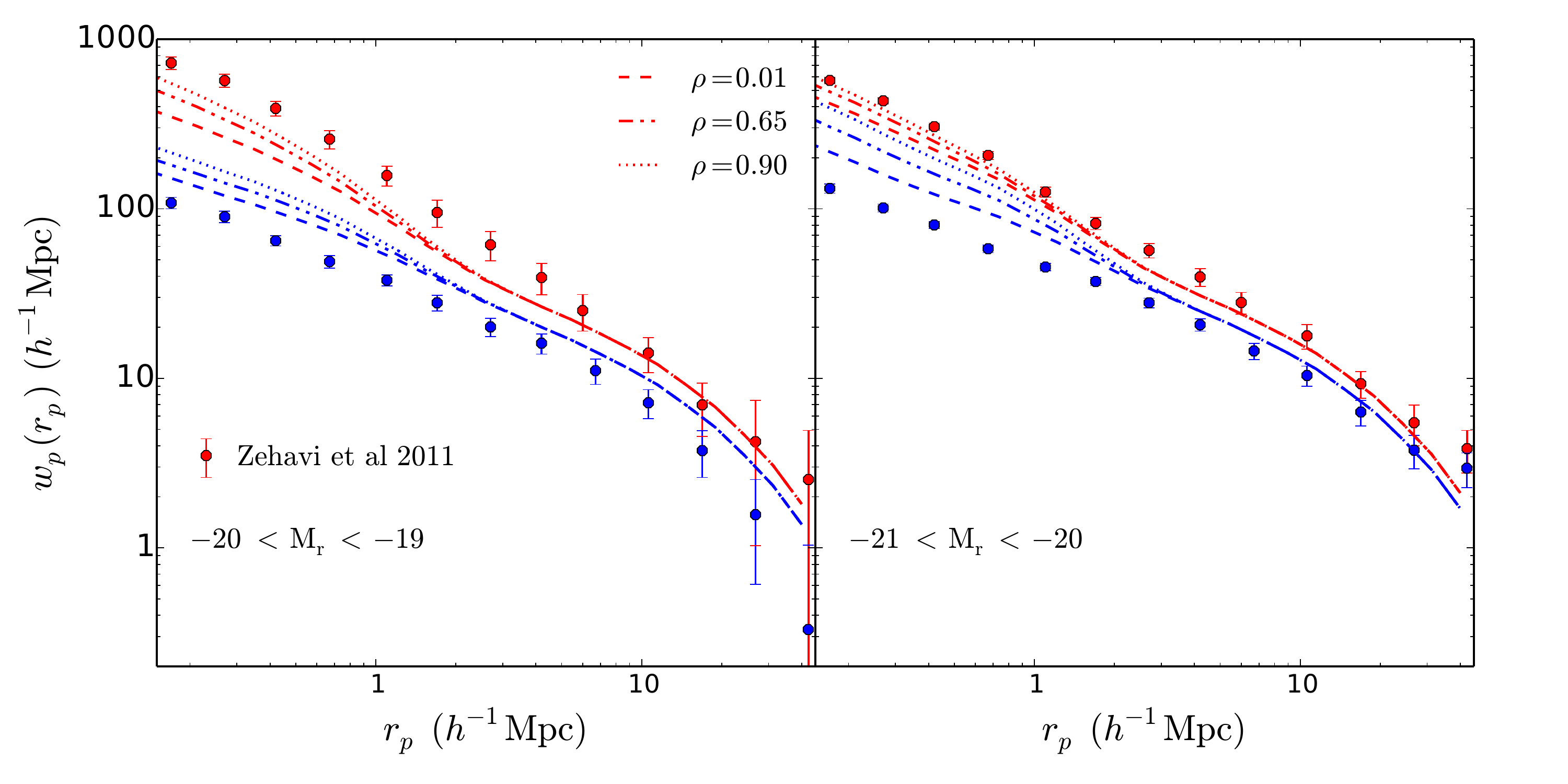}
\caption{Comparison of projected correlation function for red and blue galaxies (stronger and weaker clustering, respectively).  Points with errors show the measurements from SDSS galaxies (Z11), smooth lines show the results from our analytical model with \emph{no-2h} conformity and $p({\rm red}|{\rm sat})$ from Y07 calibration (equation~\ref{dbl-Gauss-fits}). As $\rho$ increases, the projected clustering increases for both red and blue galaxies. However, the separation between red and blue projected clustering more or less remains the same. Notice that none of the curves describe the data well.}
\label{fig:wprp_prsY07}
\end{figure*}

The definition of red and blue that we have been using so far was based on double Gaussian fits to the measured colour distributions at fixed luminosity. This is somewhat different from the definition used by Z11, which uses a luminosity dependent threshold in $g-r$ colour to separate red from blue. In Appendix~\ref{app:ccfg}, we quantify the impact of this difference on  clustering measurements, finding that 
the resulting systematic effects are subdominant compared to the other effects discussed previously. We therefore ignore this difference in what follows. For the time being, we will only consider the \emph{no-2h} model which is on firmer observational footing than the model with 2-halo conformity; we will discuss the latter towards the end of this section.

Figure~\ref{fig:wprp_prsY07} shows the projected correlation function of blue and red galaxies in SDSS (points with errors) in the two bins of luminosity mentioned above, from Tables~$9$ and $10$ of Z11. The smooth curves show the results of our \emph{no-2h} analytical model (i.e., setting $b(m,s)\to b(m)$) for different values of $\rho$ ranging from no conformity to strong conformity. While the no-conformity model clearly disagrees with the data, we see that \emph{so do the others}. This is also true at the smallest scales where, based on the results of P15, we do expect the $\rho=0.65$ model, at least, to perform reasonably well. As discussed in the previous section, the effect of increasing the conformity strength is to increase the clustering of \emph{both} red and blue galaxies (by different amounts), implying that varying $\rho$ can, at best, help describe the clustering of only one of the populations.

There are two main effects that could alleviate this disagreement at small scales, which we explore next. The first is our choice of satellite red fraction in \eqn{dbl-Gauss-fits}, which was calibrated by P15 to match measurements in the Y07 group catalog. 
The identification of satellites by the Y07 algorithm, however, is subject to systematic effects which lead to the satellite population being impure and incomplete. A recent study by \citet{campbell+15} has shown that these are $\sim20$-$30\%$ effects (see their Figure~6), and lead to a systematic  \emph{underestimation} of the satellite red fraction at fixed luminosity by about $\sim10\%$ (see their Figure~12)\footnote{This is also broadly consistent with the results of \citet{skibba09}, who compared the S09 calibration with the Y07 catalog, finding good agreement between the satellite red fractions at the bright end, with the level of agreement degrading for fainter galaxies.}. 

The second effect is our choice of using the same spatial profile for the distribution of all satellites. However, as mentioned in section~\ref{subsec:colourcorrfunc}, our model is flexible enough to allow the mean NFW concentration parameter for red galaxies to be different than blue ones. We briefly explore this possibility below.

Both of these effects will change the predicted small scale clustering of red and blue galaxies, and changing the satellite red fraction will also affect the large scale galaxy bias. At scales $0.2\lesssim r_{\rm p}/(h^{-1}{\rm Mpc})\lesssim1$, increasing the satellite red fraction $p({\rm red}|{\rm sat},L)$  will increase the 1-halo term for the red sample, while decreasing it for blue galaxies. This effect is therefore somewhat orthogonal to that of changing conformity strength, which simultaneously increases/decreases the clustering strength of both red and blue galaxies. Increasing (decreasing) the median concentration of red and blue galaxies independently will make their 1-halo clustering steeper (shallower), and can be strongly degenerate with both of the other effects. At larger scales, increasing $p({\rm red}|{\rm sat},L)$ at fixed luminosity will upweight the contribution of intermediate-to-high mass haloes (the hosts of the satellites) in the clustering of red galaxies, while downweighting it for blue galaxies. Consequently, the large scale bias of red (blue) galaxies will increase (decrease)\footnote{This is also qualitatively the same trend as induced by switching on 2-halo conformity (see section~\ref{subsubsec:2hconf}).}.

\begin{figure}
\centering
\includegraphics[width=0.45\textwidth]{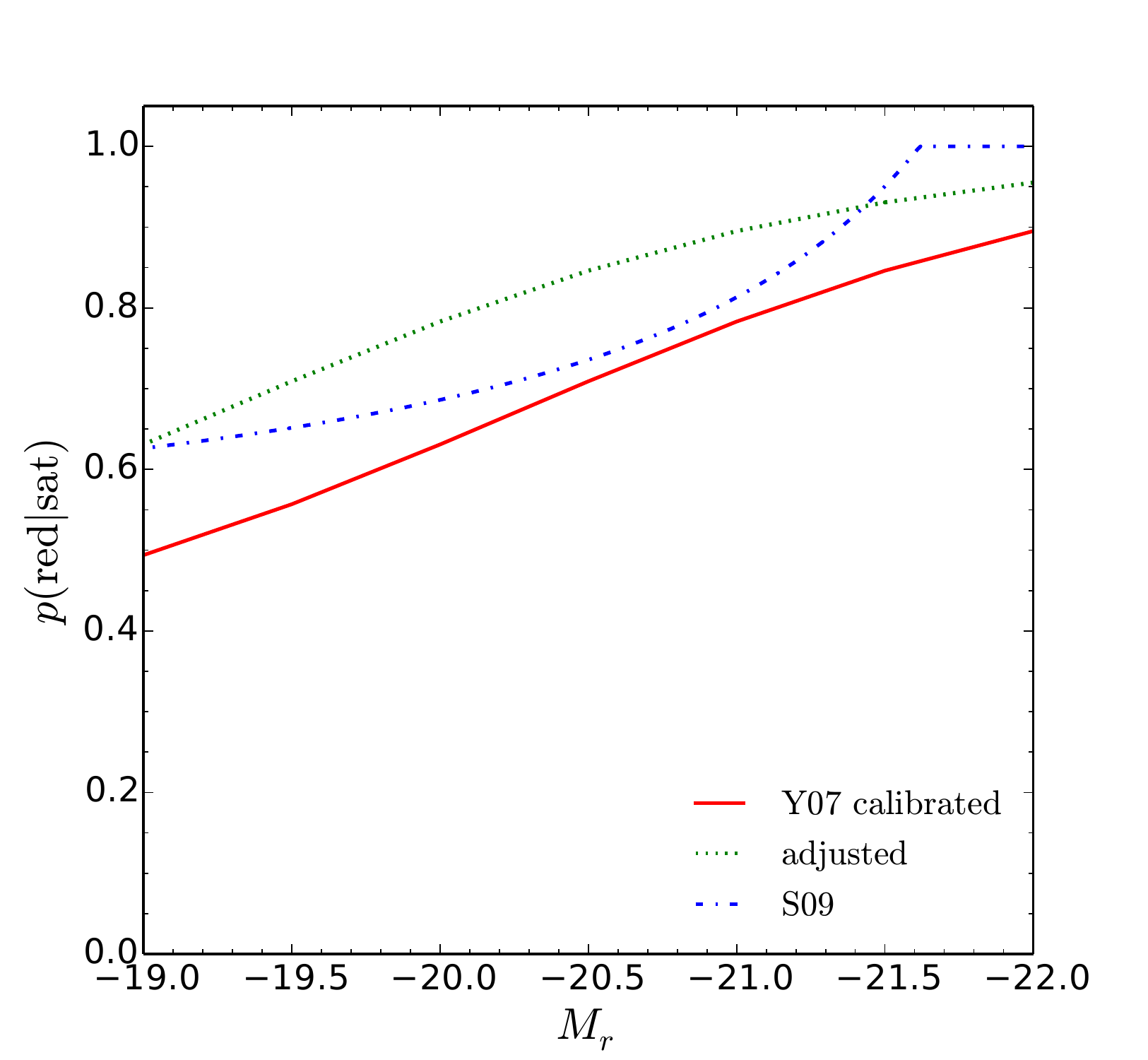}
\caption{Comparison of red fractions of satellites: solid line shows the $p({\rm red}|{\rm sat})$ from Y07 calibration (equation~\ref{dbl-Gauss-fits}), dash-dotted line represents the S09 calibration  (their equation~$8$) and the dotted one shows the $p({\rm red}|{\rm sat})$ which is {\it adjusted} to describe the projected clustering of SDSS DR7 galaxies from Z11 (equation~\ref{eq:prs-adjust}). See text for a discussion.}
\label{fig:prs}
\end{figure}

Finally, \citet{campbell+15} have shown that, despite the issues of incompleteness and impurity in the Y07 algorithm, the impact of these systematics on the measured strength of 1-halo conformity is small. In particular, the algorithm tends to introduce a spurious 1-halo conformity that can be $\sim10\%$ of the actual conformity in the sample (see their Figure~14). Consequently, we expect that setting $\rho\simeq0.6$-$0.65$ in our model, i.e., not more than $\sim10\%$ smaller than the value suggested by P15, should be a robust choice. 

\begin{figure*}
\centering
\includegraphics[width=0.9\textwidth]{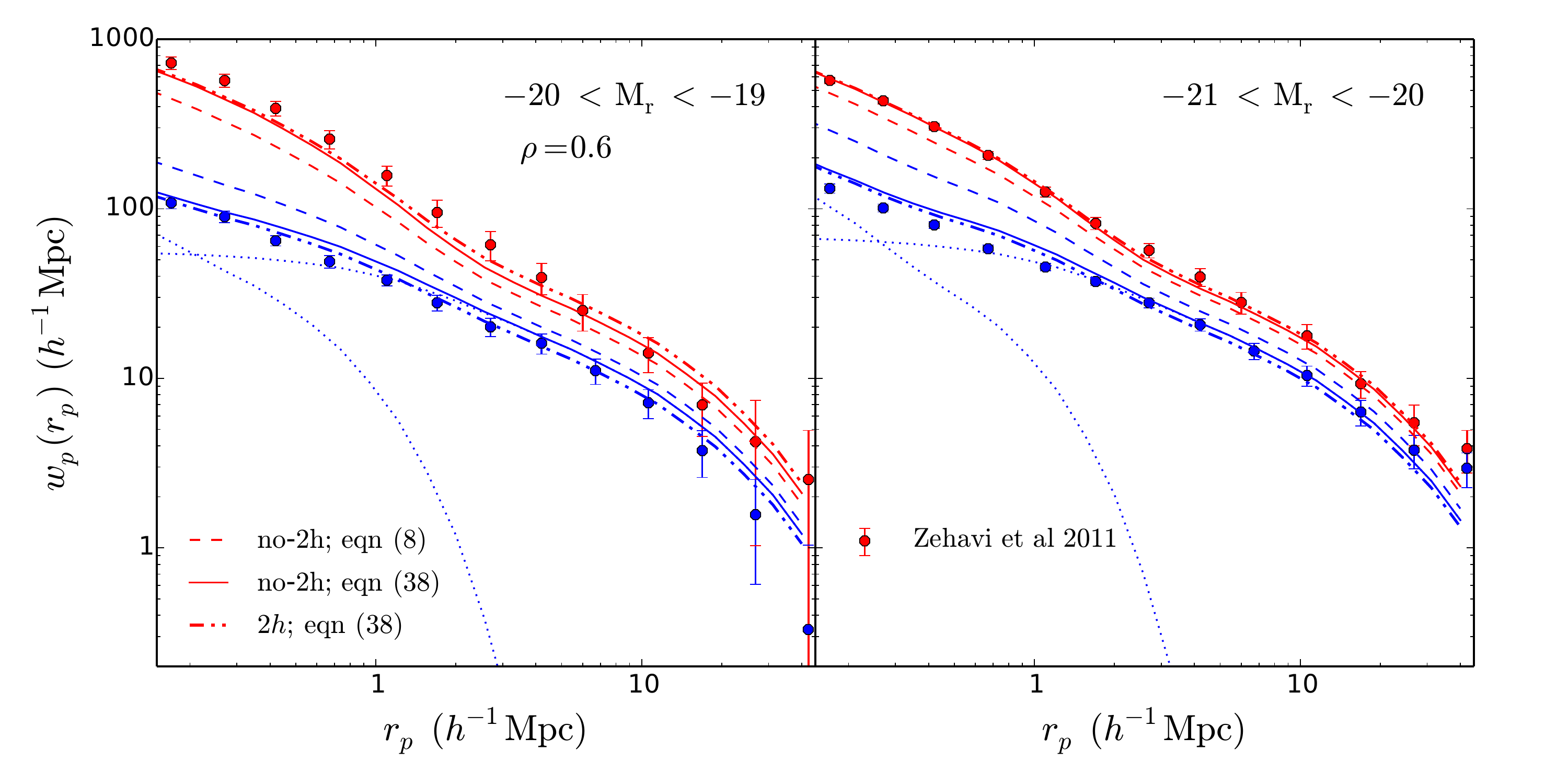}
\caption{Comparison of projected correlation function for red and blue galaxies (stronger and weaker clustering, respectively). Points with errors show the measurements from SDSS galaxies (Z11). Smooth curves show our analytical model with $\rho=0.6$. Results are shown for the \emph{no-2h} model with the Y07 calibration of $p({\rm red}|{\rm sat},L)$ (equation~\ref{dbl-Gauss-fits}; dashed curves), the \emph{no-2h} model with the `adjusted' $p({\rm red}|{\rm sat},L)$ of equation~(\ref{eq:prs-adjust}; solid curves), and the model with 2-halo conformity switched on and the adjusted $p({\rm red}|{\rm sat})$ (dot-dashed curves). Dotted curves show the decomposition of the 1-halo and 2-halo contributions to the \emph{no-2h} adjusted prediction for blue galaxies.}
\label{fig:wprp}
\end{figure*}

We have attempted to construct a model by first varying the satellite red fraction and conformity strength by modest amounts and then adjusting the red and blue concentration suitably. We have found that the blue galaxy clustering at scales $r_{\rm p}\gtrsim2\Mpch$ -- which is dominated by the 2-halo term -- is difficult to explain unless the satellite red fraction is pushed to be $\sim25\%$ larger than the Y07 calibration (equation~\ref{dbl-Gauss-fits}). In particular, the following form performs well, as we show below:
\be
p({\rm red}|{\rm sat},M_r) = 1.0 - 0.33\left[1+\tanh\left(\frac{M_r+19.25}{2.1}\right)\right]\,.
\label{eq:prs-adjust}
\ee
Figure~\ref{fig:prs} compares this `adjusted' satellite red fraction with the Y07 calibration and the S09 calibration (their equation~$8$). Further motivated by the discussion above, we set $\rho=0.6$ to describe 1-halo conformity. Figure~\ref{fig:wprp} shows the resulting predictions for projected clustering in the \emph{no-2h} model (i.e., using $b(m,s)\to b(m)$ in the 2-halo term) and using the standard NFW mean concentration as the solid curves. 
This model significantly improves upon the one that used the Y07 satellite red fraction (shown as the dashed curves) and is within $\sim20\%$ of the data at nearly all scales.

The dotted curves in Figure~\ref{fig:wprp} show the separate contributions of the 1-halo and 2-halo terms to the blue galaxy clustering in our new model. We see that the 2-halo term gives a substantial contribution at all the scales probed. Not surprisingly, then, we have found that changing the mean concentrations of red and blue satellites (by a factor of 1.7 and 0.5), respectively, of the standard value in the fainter bin) leads to some improvement, but only at the smallest scales probed. We have therefore not displayed these results.

Finally, the dot-dashed curves in Figure~\ref{fig:wprp} show the result of switching on 2-halo conformity in our new model. While there is some improvement, the overall change compared to the \emph{no-2h} model is comparable to the systematic uncertainties in our analytical approximations. We therefore conclude that there is no strong evidence of genuine 2-halo conformity in SDSS clustering data, although this inference may change if the modelling uncertainties can be brought under better control. We discuss this briefly in section~\ref{sec:concl}.

\section{Conclusions}
\label{sec:concl}
\noindent
We have presented a fully analytical Halo Model of colour-dependent galaxy clustering that incorporates the effects of galactic conformity. The model is based on our previous numerical work \citep[][P15]{pkhp15} and distinguishes between the effects of conformity at the individual group level (1-halo conformity) and potential effects due to a combination of these group-level effects with halo assembly bias (2-halo conformity). Our model ingredients include: 
\begin{itemize}
\item A calibration of the luminosity-dependent Halo Occupation Distribution (HOD). We use SDSS DR7 results from \citet{zehavi+2011}.
\item A calibration of the split between red and blue galaxies based on double-Gaussian fits to the all-galaxy colour-luminosity distribution. We use SDSS DR7 results from P15.
\item A calibration of the luminosity-dependent red fraction of satellites $p({\rm red}|{\rm sat},L)$. We discuss this below.
\item A choice of conformity strength (`group quenching efficiency') $0\leq\rho\leq1$. We explore various values, with the default being $\rho=0.65$ as suggested by P15.
\item A choice of whether or not to switch on 2-halo conformity driven by halo assembly bias. Switching this on requires a parameterisation of the concentration dependence of assembly bias (section~\ref{subsec:assemblybias} and Appendix~\ref{app:bias-comp}).
\end{itemize}
Using mock galaxy catalogs, we have demonstrated that our model correctly describes the resulting trends in colour-dependent galaxy clustering at small and large scales, for varying levels of 1-halo and 2-halo conformity, with an accuracy of $\sim10$-$20\%$. 

Rather than using the ratios of traditional correlation functions for our work, one might also think of using mark correlation functions~\citep{st04,wechsler+06,ss09}. The reason we have focussed on the ratios of correlation functions is that the effect of assembly bias on colour mark correlation functions is smaller in comparison to that on the traditional correlation functions. We can see this by comparing our results with that of \cite{zu_mandelbaum17}. We compare the percentage change of $\xi_{\rm bb}/\xi_{rr}-1$ for $\rho=0.65\,2h$ (as suggested by P15) and $\rho=0.01$ in Figure~\ref{fig:corr-conf-thresh} to the colour mark correlation functions (Figure~$8$) of \cite{zu_mandelbaum17}. The mark correlation functions only show a change of maximum $1\%$ over all stellar-mass bins at $r=6\,h^{-1}$Mpc whereas the ratios of traditional correlation functions depict the change to be between $45\%-65\%$ which is a huge difference. This motivates us to use the traditional correlation functions for this study.

Having chosen to work with traditional 2-point clustering measures, we can directly compare our model with SDSS clustering data, \emph{without} depending on any group-finder algorithm. As an application, we have used our model to assess whether or not SDSS clustering data requires the presence of 2-halo conformity; in other words, whether there are any traces of \emph{galaxy} assembly bias in the colour-dependent clustering of SDSS galaxies. This question has important ramifications for models of galaxy formation, and has gained considerable interest recently \citep{lin+16,miyatake+16,saito+16,twcm16,zu+17}. Our main findings are the following:
\begin{itemize}
\item Using the \citet[][Y07]{yang+07} result for $p({\rm red}|{\rm sat},L)$, colour-dependent clustering in SDSS \emph{cannot} be explained using \emph{any} model of conformity, including the model with no conformity (Figure~\ref{fig:wprp_prsY07}). At small scales, in particular, we see discrepancies of factors $\gtrsim2$, much larger than the expected systematics due to our various approximations. 
\item Motivated by the results of \citet{campbell+15} on systematic effects introduced by the Y07 group-finding algorithm in the satellite red fraction and conformity strength, we allowed $p({\rm red}|{\rm sat},L)$ to be free and fixed the \emph{1-halo conformity} using $\rho=0.6$, slightly smaller than that suggested by P15. We find that, upon  setting  $p({\rm red}|{\rm sat},L)$ to the expression in \eqn{eq:prs-adjust} (which is about $\sim25\%$ higher than in the Y07 catalog; Figure~\ref{fig:prs}), a model \emph{without} 2-halo conformity already gives a good description of red and blue galaxy clustering in SDSS (solid curves in Figure~\ref{fig:wprp}).
\item Switching on 2-halo conformity does not lead to significant improvement between the model and data at large scales (dotted curves in Figure~\ref{fig:wprp}). Given the $10$-$20\%$ systematic uncertainty in our model, we therefore do not find strong evidence of genuine 2-halo conformity driven by halo assembly bias in SDSS clustering data. 
\end{itemize}

We have presented our main results for luminosity, $M_r<-21.0$. This is worth mentioning here because the SDSS luminosities are likely to be affected by the systematics effects for $M_r<-21.3$~\citep{bernardi+13}. Since we do not explore regimes for luminosities higher than $M_r<-21.0$,  our results are not effected by these systematics.

The systematic uncertainties arising from our analytical approximations, particularly in the notorious 1-halo to 2-halo transition regime, could be reduced by taking inputs from $N$-body simulations. E.g., \citet{zg16} used tabulated measurements of (sub)halo correlation functions and dark matter profiles in high resolution $N$-body simulations to improve the accuracy of the HOD models. Their techniques can, in principle, be adapted to include conformity in a straightforward manner. Our results could also be strengthened by systematically allowing the conformity strength, satellite red fraction and median red and blue spatial profiles to vary simultaneously, using Markov Chain Monte Carlo techniques together with appropriate priors on these quantities, as well as the full error covariance matrix of the clustering measurements. We will take up these exercises in future work. 

\section*{Acknowledgments}
\noindent
The research of AP is supported by the Associateship Scheme of ICTP, Trieste and the Ramanujan Fellowship awarded by the Department of Science and Technology, Government of India.

\bibliography{allrefs}

\appendix

\section{testing the calibration of halo assembly bias}
\label{app:bias-comp}

\begin{figure*}
\centering
\includegraphics[width=0.95\textwidth]{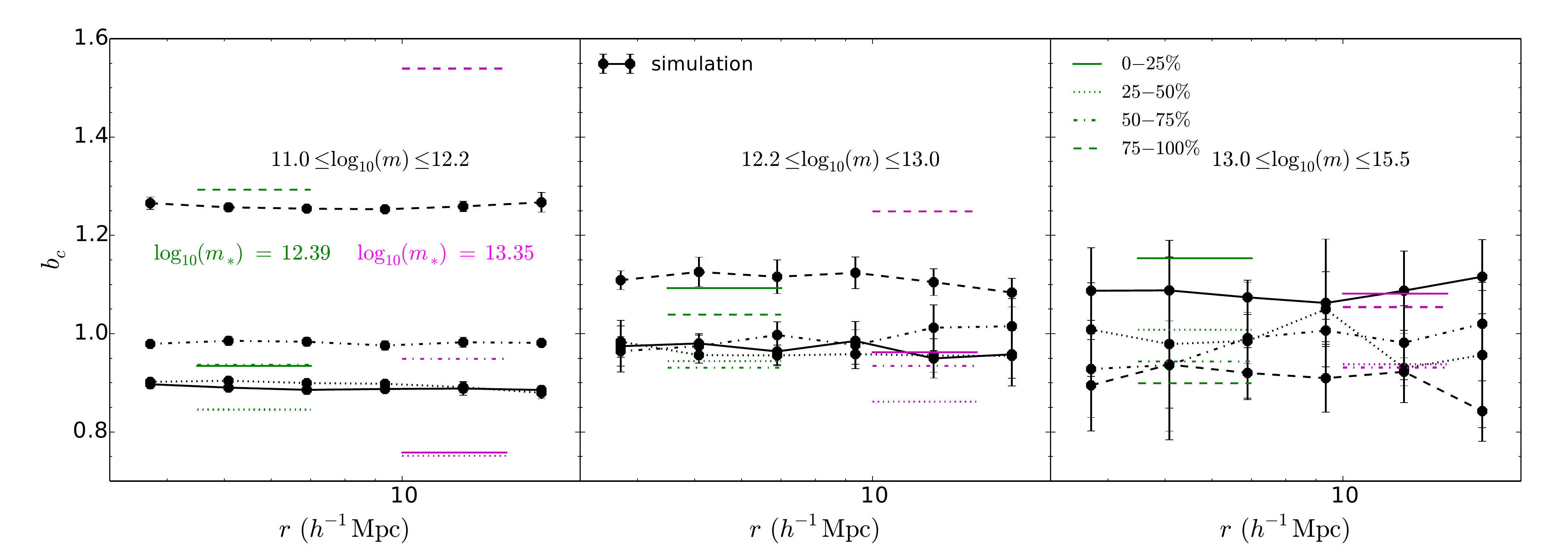}
\caption{Comparison of halo assembly bias in simulation and analytical model: points with errors shows the simulation data and the smooth lines show the halo assembly bias from the fitting function (equation~\ref{eq:wechr-fit}). Each panel corresponds to a mass bin and is divided into two parts, with $b_c$ from fitting function plotted using $m_\ast = 2.47*10^{12}h^{-1}{\rm M}_{\odot}$ in the left part and using $m_\ast = 2.22*10^{13}h^{-1}{\rm M}_{\odot}$ in the right part. The different line-styles correspond to the concentration quartiles as mentioned in the third panel. }
\label{fig:bias-comp}
\end{figure*}
\noindent
In this Appendix, we compare the assembly bias fitting formula of \citet[][hereafter, W06]{wechsler+06} with measurements of the concentration dependence of the correlation function of haloes in our $N$-body simulations. W06 parametrised the concentration and mass dependence of linear halo bias as follows:
\begin{align}
b(m,s)\,=\,b(m)\,b_{c}(\tilde m,s),
\end{align}
where $b(m)$ is the bias of all haloes of mass $m$, $s$ is the standardized halo concentration defined in \eqn{eq:sdef} and $\tilde m = m/m_{\rm inv}$, where $m_{\rm inv}$ is the mass scale where the assembly bias trend with concentration inverts. W06 set $m_{\rm inv}=m_\ast(z)$, where $m_\ast(z)$ satisfies $\nu(m_\ast, z) = 1$. The quantity $b_c(\tilde{m},s)$ is defined to be the relative bias of haloes as a function of $s$, and was estimated by W06 as the ratio of the correlation function of haloes, $\xi(r|\tilde{m},s)$ of fixed $\tilde{m}$ and fixed $s$ relative to the correlation function of all haloes, $\xi(r|\tilde{m})$ of fixed $\tilde{m}$, 
\begin{align}
b^2_c(\tilde{m},s) & \equiv \xi(r|\tilde{m},s)/\xi(r|\tilde{m}),
\end{align}
taking an average over separations from $5\le r/(h^{-1}{\rm Mpc})\le10$. A fitting function to the simulation data has been provided by W06 and is of the form 
\begin{align}
b_c(\tilde{m},s) &= p(\tilde{m}) + q(\tilde{m}) s + 1.16 [1-p(\tilde{m})] s^2, \label{eq:wechr-fit}
\end{align}
where
\begin{align}
p(\tilde{m}) &=0.95+0.042 \ln(\tilde{m}^{0.33}), \notag\\
q(\tilde{m}) &= 0.1 - \frac{0.22[\tilde{m}^{0.33} + \ln(\tilde{m})^{0.33}]}{(1+\tilde{m}^{0.33})}. \notag
\end{align}
We next investigate the accuracy with which this fitting function describes assembly bias in our simulations. Before we do so, it is useful to digress a bit and discuss the reason why the inversion scale should be associated with $\nu=1$.

Although Figure~6 of W06 is consistent with $m_{\rm inv}=m_\ast$, the statistical errors in their measurements leave room for a variation of a factor of a few, which is worth exploring, as we show below. In fact, the typical rationale behind $m_\ast(z)$ satisfying $\nu(m_\ast,z)=1$ is that this mass scale is a `characteristic' scale representing the peak of the mass \emph{fraction}: $\nu f(\nu) = (m/\bar\rho)n(m)|\der\ln\nu/\der\ln m|$. For the original Press-Schechter mass function, $\nu f(\nu)\propto\nu\,{\rm e}^{-\nu^2/2}$ \citep{ps74} which peaks at $\nu_\ast=1$, thus justifying the definition of $m_\ast$. More modern calibrations of the mass function show that the actual peak occurs at substantially higher masses. For our cosmology, at $z=0$, $m_\ast = 2.47\times10^{12}\Mh\equiv m_1$ when using the Press-Schechter definition $\nu=1$, while $m_\ast = 2.22\times10^{13}\Mh\equiv m_2$ when using the peak of the mass fraction derived from the \citet{Tinker08} halo mass function. If the inversion scale is indeed tied to the characteristic mass scale, then the more natural value would seem to be the larger one. Purely from first principles, however, there is no reason to expect that \emph{either} of these scales is associated with the inversion, since there is no robust analytical model of assembly bias as yet \citep[see][for some attempts]{dalal+08,cs13}.

Figure~\ref{fig:bias-comp} shows the comparison of $b_c(\tilde{m},s)$ as obtained from our simulation data (points with errors) to the one obtained from the fitting function (horizontal line segments) mentioned above, for three mass bins (one in each panel). The points show the $b_c$ in six bins of separation $3\lesssim r/(h^{-1}{\rm Mpc})\lesssim20$. Each panel is divided into two parts, with $b_c$ from fitting function using $m_{\rm inv} = m_1$ in the left part (coloured green) and using $m_{\rm inv} = m_2$ in the right part (coloured magenta). The different line styles corresponds to the concentration quartiles as indicated in the legend. 

We see that neither of the fit values describe the simulation data well. However, there is a trend. In the low mass bin (left panel), all constant lines follow the trend of that of simulation data. In other words, the highest concentration quartile corresponds to the highest bias and the lowest one to the smallest bias. In the middle panel, the trend is still same for the simulation data, however, the inversion of bias takes place for the fitting function with $m_1$ because now $m>m_1$. Thus, the trend for the fit with  $m_1$ starts inverting, the upper most quartile becomes less biased and the lowest quartile attains the highest bias. For $m_2$, the inversion of bias has not happened yet ($m<m_2$) and the trend is very similar to that of simulation data. In the last mass bin, the inversion of bias has taken place completely for simulation data whereas the inversion is taking place for $m_2$. We conclude that the inversion mass scale for the simulation data is somewhat less than $m_2$ but it is certainly higher than $m_1$.

As the trend in the simulation data is closer to that obtained using \eqn{eq:wechr-fit} with $m_2$ rather than $m_1$ as W06 suggested, we use $m_2$ to calculate the assembly bias for our analytical model. We have also checked that using $m_1$ instead of $m_2$ substantially worsens the agreement between our analytical model and mock data in Figure~\ref{fig:corr-conf-thresh}.
We note however, that this is only an approximate fix; studies such as ours require a more accurate calibration of assembly bias. Fortunately, our main conclusions are unaffected by the systematic uncertainty in the calibration above.

\section{Effect of definition of galaxy colour on $\xi(r)$}
\label{app:ccfg}
\noindent
In this appendix, we investigate the effect of using different definitions of `red' and `blue' for galaxy classification on the correlation function. The distribution of $g-r$ colour for the SDSS galaxies are well modelled  as the sum of two Gaussian components \citep{baldry+04,ss09}. The relative fraction of these two components was used by S09 to define a stochastic variable (a `red flag') which would determine the component from which the galaxy's $g-r$ colour is drawn. A galaxy is then `red' if it colour was drawn from the `red' Gaussian, and similarly for `blue'.
Observationally, however, it is more convenient to simply impose a hard cut in $g-r$, labelling galaxies of luminosity $M_r$ as `red' if their $g-r$ colour exceeds
\begin{align}
(g-r)_{\rm cut} = 0.8 - 0.03 (M_r+20)\,,
\label{eq:colorcut}
\end{align}
and as `blue' otherwise. The difference between these definitions can introduce some uncertainty when comparing model predictions to observations.

\begin{figure}
\centering
\includegraphics[width=0.5\textwidth]{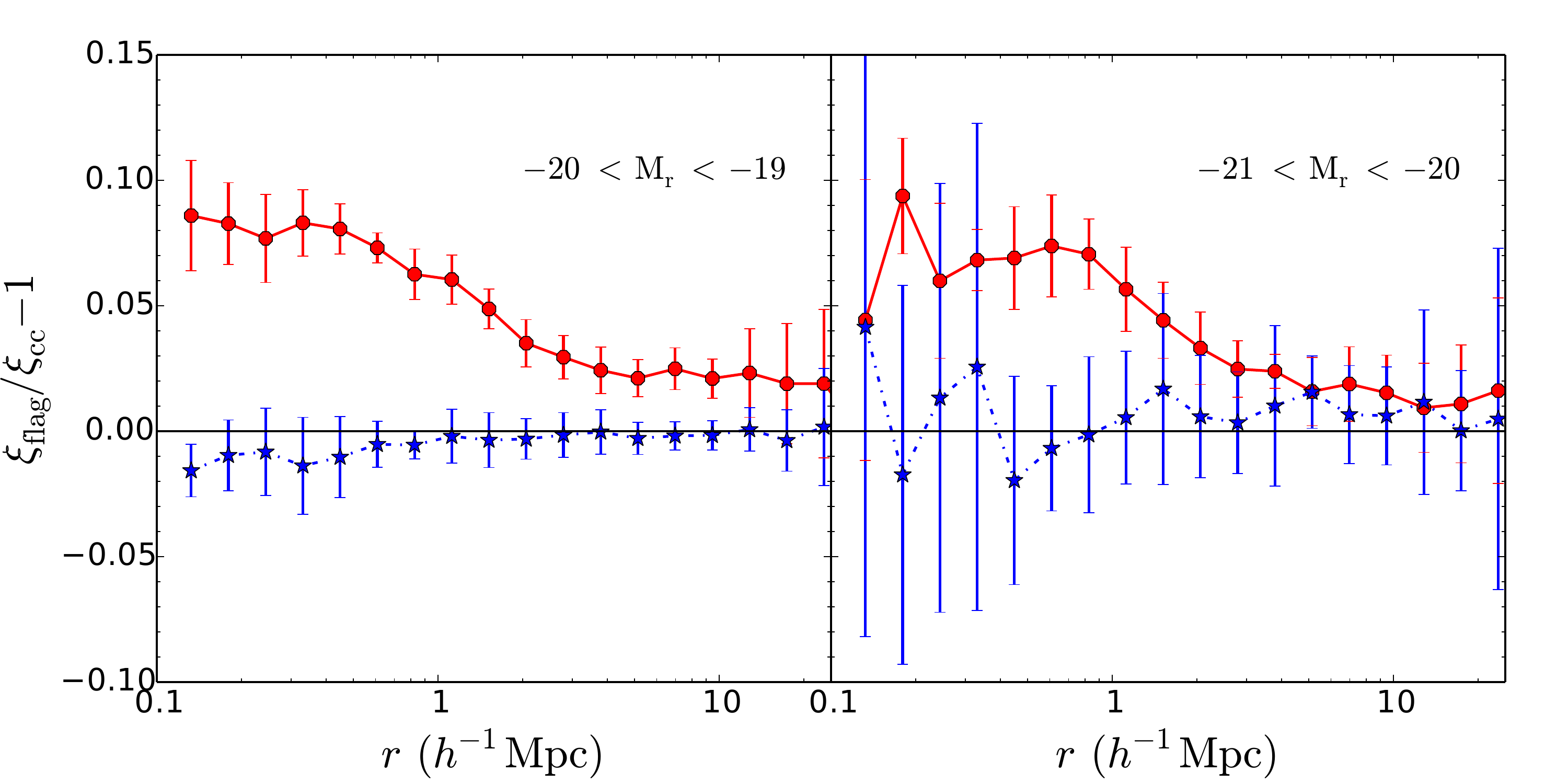}
\caption{Comparison of the correlation function based on `cut' and `flag' definitions of the colour of galaxy:  Circles (stars) show the excess of the correlation function based on the `flag' definition of the colour to that obtained from the `cut' definition of the colour for red (blue) galaxies. There is a $\sim10\%$ difference on the small scale measurements and $2$-$3\%$ on large scales for the red clustering while blue clustering shows little change at all scales.}
\label{fig:xi-cc-fg}
\end{figure}

Since our analytical model is built upon the S09 algorithm, it inherently defines galaxies as `red' or `blue' using the `flag' definition. The mock catalogs we use, on the other hand, contain both, the value of the red flag for each object, as well as its actual $g-r$ colour. In the main text, we always compared our analytical results to corresponding mock measurements wherein galaxies were split by their red flag values. Figure~\ref{fig:xi-cc-fg} explores the systematic error introduced in the correlation function if one were to split galaxies by the colour cut \eqref{eq:colorcut} instead. The Figure shows the excess of the correlation function based on the `flag' definition to the one based on `colour-cut' definition for red (circles) and blue (stars) galaxies. We see that the difference in the red clustering based on both definitions is about $\sim 10\%$ at small scales and  $\sim 2$-$3\%$ on large scales whereas the blue clustering does not change much with change in the definition of the colour of galaxy. Since this is smaller than the other systematic effects in our analytical model at nearly all scales, we ignore this difference when comparing with SDSS observations in the main text.

\label{lastpage}

\end{document}